\documentclass[aps,prd,11pt,superscriptaddress,notitlepage,longbibliography,nofootinbib,tightenlines,groupedaddress]{revtex4-1}
\usepackage{amsmath,amssymb,amsfonts}
\usepackage{graphicx}
\usepackage{color}
\usepackage{tikz}
\usepackage{booktabs}
\usepackage{dsfont}
\usepackage[pdftex]{hyperref}
\hypersetup{colorlinks=true, linkcolor=darkred, citecolor=blue, linktoc=page}
\definecolor{darkred}{rgb}{0.8,0.1,0.1}

\newcommand{\N}[1]{\ensuremath{\mathcal N\,{=}\,#1}}
\newcommand{\Nc}{N_{\rm c}}

\makeatletter\def\l@subsubsection#1#2{}%
\makeatother

\newcommand{\nocontentsline}[3]{}
\newcommand{\tocless}[2]{\bgroup\let\addcontentsline=\nocontentsline#1{#2}\egroup}
\def\Im{\mathop{\rm Im}}
\def\Re{\mathop{\rm Re}}
\def\ie{{\it i.e.}}

\def\q{{\bf q}}
\def\x{{\bf x}}
\def\qq{q}
\def\ww{\omega}
\def\uh{u_{\rm h}}
\def\O{{\mathcal O}}
\def\half{\tfrac{1}{2}}
\def\h{\mathfrak {h}}

\begin{document}

\title{Damping of hard excitations in strongly coupled \texorpdfstring{$\N4$}{N=4} plasma}

\author{John F.~Fuini III}
\email{fuini@uw.edu}
\author{Christoph F.~Uhlemann}
\email{uhlemann@uw.edu}
\author{Laurence G.~Yaffe}
\email{yaffe@phys.washington.edu}
\affiliation{Department of Physics, University of Washington, Seattle, WA 98195-1560, USA}

\begin{abstract}
    The damping of high momentum excitations in strongly coupled maximally
    supersymmetric Yang-Mills plasma is studied.
    Previous calculations of the asymptotic behavior of the quasinormal mode
    spectrum are extended and clarified.
    We confirm that subleading corrections to the lightlike dispersion relation
    $\omega(\q) = |\q|$ have a universal $|\q|^{-1/3}$ form.
    Sufficiently narrow, weak planar shocks may be viewed as coherent
    superpositions of short wavelength quasinormal modes.
    The attenuation and evolution in profile of 
    narrow planar shocks are examined as an application of our results.
\end{abstract}

\maketitle
\tableofcontents

\section{Introduction}

In a strongly coupled Yang-Mills plasma, such as that of
maximally supersymmetric Yang-Mills ($\N4$ SYM) theory,
the typical time scale for relaxation of non-hydrodynamic perturbations
is set by the inverse temperature $T^{-1}$.
In a dual holographic description,
this scale may be interpreted as the characteristic
gravitational infall time for perturbations falling through the
horizon of black brane geometries which describe near-equilibrium states
\cite{Chesler:2008hg,Bhattacharyya:2009uu}.

However, even in the strong coupling limit,
sufficiently high momentum excitations are only weakly damped.
This may, for example, be seen in the large wavenumber asymptotics
of the quasinormal mode (QNM) spectrum.
At zero temperature in $\N4$ SYM,
Fourier transformed two-point correlation functions,
viewed as functions of frequency $\omega$ at fixed wavenumber $\q$,
have branch cuts starting at the lightcone, $\omega = \pm |\q|$.%
\footnote
    {%
    Throughout this paper, we consider $\N{4}$ SYM theory on $\mathbb R^4$,
    or the dual gravitational theory on the Poincar\'e patch of the
    AdS$_5$-Schwarzschild geometry.
    }
At non-zero temperature, and in the $N \to \infty$ limit,
this branch cut breaks up into a closely spaced series of poles
at locations $\omega = \{ \omega_n^\pm (\q) \}$ known as quasinormal
mode frequencies 
\cite{Starinets:2002br,Nunez:2003eq,Kovtun:2005ev}.
Festuccia and Liu \cite{Festuccia:2008zx}
studied the large-$\q$ asymptotics of the quasinormal
mode spectrum for scalar perturbations
(or helicity $\pm 2$ stress-energy perturbations) and found that
as $|\q| \to \infty$,
\begin{equation}
    \omega_n^\pm(\q)/|\q| \sim
    \pm \left[ 1 + c_n \, e^{\mp i \pi/3} \, (\pi T/|\q|)^{4/3} \right],
\label{eq:leadingform}
\end{equation}
with real ``spectral deviation'' coefficients $\{ c_n \}$ which are discussed below.
The small imaginary part (relative to the real part),
$\Im \omega_n^\pm(\q)/ \Re \omega_n^\pm(\q) = \O(T/|\q|)^{4/3}$,
demonstrates the weak damping for $|\q| \gg T$,
and shows that these high energy, short wavelength excitations may,
in some respects, be regarded as quasiparticles, \ie,
excitations whose mean free path is much longer than their
de Broglie wavelength.
However, because $|\q| \gg T$, these are highly athermal excitations
which are exponentially rare in equilibrium.
Moreover, because the spacing in energy between successive
quasinormal modes is comparable to their width,
$|\omega_{n+1}^\pm(\q) - \omega_n^\pm(\q)| \sim |\Im \omega_n^\pm(\q)|$,
the spectral densities of correlation functions, at large $\q$,
do not have distinct narrow peaks in frequency associated with each
quasinormal mode;
instead the contributions of multiple QNMs merge to produce
a slowly varying spectral density \cite{Steineder:2013ana}.

The weak damping of high $\q$ excitations may also be seen in
the behavior of planar shocks.%
\footnote
    {%
    By ``planar shock'' we mean a state with an energy
    density distribution resembling a uniform infinite planar sheet,
    with a longitudinal profile characterized by some width $w$,
    and propagating in a direction normal to the sheet.
    }
At zero temperature, planar shocks propagate at the speed of light
without dispersion or attenuation.
At non-zero but low temperatures
(small compared to the inverse width of the shock),
the shock experiences weak thermal drag \cite{Chesler:2010bi}.
This slowly attenuates the amplitude of the shock
and introduces dispersion, but this weak damping vanishes as $T \to 0$.

In this paper, we study the damping of high $\q$ excitations in
$\N4$ SYM theory in greater detail.
In section \ref{sec:QNM} we perform our own WKB analysis 
of the large-$\q$ asymptotics of helicity $\pm2$ quasinormal mode frequencies.
We confirm the relative $|\q|^{-4/3}$ form (\ref{eq:leadingform})
of the leading corrections to a lightlike dispersion relation,
with a universal $\mp\pi/3$ phase.
However, we find values for the coefficients $\{ c_n \}$ of these corrections
which disagree in two respects with the result
given in ref.~\cite{Festuccia:2008zx},
which was
\begin{equation}
    c_n = K_{\rm FL} \, (2n + 1)^{4/3} \,,
    \qquad n = 1, 2, \cdots,
    \qquad \mbox{[Festuccia \& Liu]}\,,
\label{eq:FLresult}
\end{equation}
with
$
    K_{\rm FL} = [\sqrt\pi \, \Gamma\big(\tfrac 74\big)/\Gamma\big(\tfrac 14\big)]^{4/3}
    = 0.344127\cdots
$.
The $(2n+1)^{4/3}$ dependence on mode number is asymptotically correct for
high-lying modes, but is not accurate for low order modes.
Moreover,
the coefficient $K_{\rm FL}$ differs by a factor of $2^{5/3}$
from the correct value in the large order asymptotic form,
\begin{subequations}\label{eq:cn helicity2}
\begin{align}
    c_n &\sim K \, (2n + 1)^{4/3} \,,
    \qquad n \gg 1 \,,
\\
\noalign{\hbox{with}}
    K &=
    \half [\sqrt\pi \,\Gamma\big(\tfrac 74\big)/\Gamma\big(\tfrac 54\big)]^{4/3}
    = 1.092535\cdots \,.
\end{align}
\end{subequations}
The need for a $2^{5/3}$ correction factor 
in the value of the coefficient $K_{\rm FL}$ for AdS$_5$ black holes
was noted earlier in ref.~\cite{Morgan:2009vg},%
\footnote
    {%
    We thank G.~Festuccia for making us aware of this reference.
    }
but the inaccuracy of the estimate (\ref{eq:cn helicity2}) for low order
modes seems not to have been previously appreciated.

Complementary numerical results confirming the WKB analysis,
examining the approach to the asymptotic regime,
and studying helicity 0 and $\pm1$ modes in addition to helicity $\pm2$,
are presented in section \ref{sec:numerics}.
We calculate accurate results for the lowest
fifteen quasinormal modes in each helicity channel
for wavevectors up to $|\q|/(\pi T) = 160$.
This extends previous results given in ref.~\cite{Kovtun:2005ev}.
For helicity $\pm 2$ perturbations, comparison of the numerical results with the WKB
asymptotics clearly confirms the validity of the asymptotic analysis and
shows that for low order modes the large-$\q$ asymptotic form (\ref{eq:leadingform})
becomes a good approximation starting at modest wavenumbers of a few times $\pi T$.
For helicity $\pm1$ and 0 perturbations (which satisfy significantly more complicated equations)
we have not performed a full WKB asymptotic analysis.
However, our numerical results for these helicities very clearly support the assertion
that the asymptotic form (\ref{eq:leadingform}) is equally valid for these perturbations.
Moreover, our extrapolated numerical values for the first fifteen spectral deviation coefficients
$c_n$ strongly suggest that in these helicity channels the large order asymptotic form is
\begin{equation}
    c_n \sim \begin{cases}
	K \, (2n)^{4/3} \,, & \mbox{helicity } {\pm}1; \\
	K \, (2n-1)^{4/3} \,, & \mbox{helicity } 0,
    \end{cases}
\label{eq:cn helicity10}
\end{equation}
with exactly the same prefactor
$ K = \half [\sqrt\pi \, \Gamma\big(\tfrac 74\big)/\Gamma\big(\tfrac 54\big)]^{4/3} $
as for helicity $\pm2$.

As an application of our results, we discuss the propagation of
narrow planar shocks in section \ref{sec:narrow-shocks}.
A sufficiently weak shock may be viewed as a coherent superposition of 
quasinormal modes.
As noted above,
as the shock moves through the dispersive $\N4$ SYM plasma at temperature $T$
it experiences friction;
the maximum amplitude will decrease and the longitudinal energy density
profile will evolve.
We specifically study narrow planar shocks
whose quasinormal mode spectrum is dominated by wavevectors large compared
to $T$ and discuss characteristic features of the resulting evolution.
The final section \ref{sec:discussion} contains a few concluding remarks.
Appendix \ref{app:table} presents tabular data for QNM frequencies.
Three subsequent appendices provide details on the numerical analysis, 
transformation to infalling coordinates,
and the large wavevector, large order asymptotic analysis.

\section{Quasinormal mode frequencies: large-\texorpdfstring{$\q$}{q} asymptotics}\label{sec:QNM}

We wish to study the dynamics of linearized perturbations on the background geometry
of an AdS$_5$ black brane,
which is dual to the thermal equilibrium state (at vanishing chemical potentials)
of $\N4$ SYM theory.
We find it convenient to use infalling coordinates $(t,\x,u)$ with
$\x \equiv (x_1,x_2,x_3)$ denoting ordinary spatial coordinates and $u$ an
(inverted) bulk radial coordinate.
Choosing to set the AdS curvature scale $L$ equal to unity,
the metric reads
\begin{equation}\label{eqn:metric}
    g =
    u^{-2}
    \left[-dt \otimes du-du \otimes dt-\left(1-m\, u^4\right)dt^2+d\x^2 \,\right].
\end{equation}
The conformal boundary is at $u=0$ and the horizon lies at $\uh \equiv (\pi T)^{-1}$,
with
\begin{equation}
    m \equiv \uh^{-4} = (\pi T)^4 \,.
\end{equation}

The metric is translationally invariant in the Minkowski directions $(t,\x)$.
Hence, it is natural to Fourier decompose the dependence of perturbations on these
directions and, for non-zero wavevectors,
to classify according to the helicity of the perturbation
under the SO(2) little group \cite{Policastro:2002se}.
In this section we focus, for simplicity, on helicity $\pm2$ perturbations.
Choosing the wavevector $\q$ to lie along the $x_3$-direction
(with magnitude $q$),
we consider a metric perturbation of the form
\begin{equation}\label{eqn:pert-ansatz}
    \delta g
    =
    u^2 \, h(u) \, e^{i(q x_3-\omega t)} 
    \left(dx_1\otimes dx_2+ dx_2 \otimes dx_1\right) ,
\end{equation}
with $h$ an undetermined function of $u$.
Factoring out two powers of $u$, as shown, is convenient
as the appropriate boundary condition for $h$ at $u = 0$ then becomes just regularity.
Similarly, because our infalling coordinates are non-singular across the future
horizon, ingoing boundary conditions at the horizon correspond to $h$ also remaining
regular at $u=\uh$.

With this choice of perturbation,
the only non-trivial part of the linearized Einstein's equations is the $xy$ component,
and the resulting equation reads
\begin{align}\label{eqn:radial-eq-0}
    h''
    + \frac{5-9\, m u^4+2 i u \omega}{u \, (1-m u^4)} \, h'
    - \frac{q^2 u+16\, m u^3-5 i \omega}{u \, (1-m u^4)} \, h=0 \,.
\end{align}
Henceforth, it is convenient to choose units such that $m = 1$
(or equivalently, to measure $\omega$ and $q$ in units of $\pi T$),
so that the helicity $\pm2$ perturbation equation becomes%
\footnote
    {%
    Factors of $\pi T$ can always be reinstated by rescaling
    $
	\omega \to \omega/(\pi T)
    $
    and
    $
	q \to q/(\pi T)
    $,
    along with
    $
	u \to u \, \pi T
    $.
    }
\begin{align}\label{eqn:radial-eq}
    h''
    +\frac{5-9 u^4+2 i u \omega}{u\,(1- u^4)} \, h'
    -\frac{q^2 u+16 u^3-5 i \omega}{u\,(1- u^4)} \, h =0 \,.
\end{align}
It will also prove convenient to 
denote the frequency to wavevector ratio by
\begin{align}\label{eqn:s-def}
    s \equiv \omega/q \,.
\end{align}
This ratio will be complex and wavenumber dependent [\ie, $s=s(q)$],
although this dependence will not always be indicated explicitly.
From the quasinormal mode equation (\ref{eqn:radial-eq})
it is apparent that if $h(u)$ is a solution with frequency $\omega$
then $h(u)^*$ is also a solution with frequency $-\omega^*$,
showing that quasinormal mode frequencies (which are not pure imaginary)
come in pairs with opposite real parts.
Hence, it is sufficient to focus on solutions with $\Re s \ge 0$.

One may eliminate first derivative terms in the helicity $\pm2$ equation
(\ref{eqn:radial-eq}) by suitably redefining the radial function.
Let 
\begin{equation}\label{eqn:radial-redef}
    h(u)=
    e^{-i \omega f(z)} \, z^{-3/2} \, (1{-}z^2)^{-1/2} \, \hat h(z) \,,
\end{equation}
with
\begin{equation}
    f(z) \equiv \tfrac 12 \left[\tan^{-1}(\sqrt{z})+\tanh^{-1}(\sqrt{z}) \right]
\end{equation}
and $z \equiv u^2$.
Then $\hat h (z)$ satisfies a zero-energy Schr\"odinger equation,
\begin{equation}\label{eqn:radial-eq-for-WKB}
    -\hat h'' + V(z) \, \hat h = 0 \,,
\end{equation}
with
\begin{equation}
    V(z)
    \equiv
    q^2 \, \frac{1-s^2-z^2}{4 z \left(1{-}z^2\right)^2}
    + \frac{3 - 6 z^2 - z^4}{4 z^2 \left(1{-}z^2\right)^2} \,.
\end{equation}
Boundary conditions on $\hat h$
(corresponding to regularity of $h$ at horizon and boundary) are
\begin{subequations}\label{eqn:hhat-bcs}
\begin{align}
    \hat h(z) 
    &=\O\big(z^{3/2}\big) \,,
    &&\mbox{as $z \to 0$} \,;
\label{eqn:hhat-bc}
\\
    \hat h(z) 
    &=\O\big((1{-}z)^{1/2-i\omega/4}\big) \,,
    &&\mbox{as $z \to 1^-$} \,.
\label{eqn:hhat-bcB}
\end{align}
\end{subequations}
The six singular points of eq.~(\ref{eqn:radial-eq})
(at $u=0$, $u=\infty$, and $u^4=1$)
are reduced to four in eq.~(\ref{eqn:radial-eq-for-WKB}): $z=0$, $z=\infty$, and $z= \pm 1$.
The resulting equation (\ref{eqn:radial-eq-for-WKB}) is thus of the Heun type.

\subsection{Leading behavior}

As mentioned earlier,
in the large $q$ (or low temperature) limit,
where the spatial wavevector is arbitrarily large compared to $\pi T$, 
quasinormal mode frequencies should approach the zero-temperature branch points at
$\omega^2 = q^2$.
To demonstrate that this is indeed the case,
we insert an ansatz for the asymptotic behavior of the ratio $s = \omega/q$,
\begin{equation}\label{eqn:s-exp}
    s(q) = s_0 + s_\alpha(q) \,q^{-\alpha} \,,
\end{equation}
with exponent $\alpha>0$ and the ``dispersive correction'' 
$s_\alpha(q)$ a smooth function of $q$ which approaches a finite non-zero limit,
\begin{equation}
    s_\alpha^\infty \equiv \lim_{q\to\infty} s_\alpha(q) \,,
\end{equation}
with corrections vanishing as an inverse power of $q$.

First,
to show that $s_0$ must equal $\pm 1$, we make a proof by contradiction:
assume that $s_0^2\neq 1$ and demonstrate that there are no solutions.
Eq.~(\ref{eqn:radial-eq-for-WKB}) becomes
\begin{equation}\label{eqn:radial-eq-for-WKB2}
    q^{-2} \, \hat h''
    =
    \left(
	Q_0
	+ \qq^{-\alpha} \, Q_\alpha
	+ \qq^{-2} \, Q_2
	+ \qq^{-2\alpha} \, Q_{2\alpha}
    \right) \hat h \,,
\end{equation}
where
\begin{align}\label{eqn:Q-gen}
 Q_0(z)&\equiv\frac{1 - s_0^2 - z^2}{4 z \left(1 {-} z^2\right)^2}\,,&
 Q_\alpha(z)&\equiv\frac{-s_0 \,s_\alpha(q)}{2 z \left(1 {-} z^2\right)^2}\,,&
 Q_2(z)&\equiv\frac{3 - 6 z^2 - z^4}{4 z^2\left(1 {-} z^2 \right)^2}\,,&
 Q_{2\alpha}(z)&\equiv\frac{-s_\alpha(q)^2}{4 z \left(1 {-} z^2\right)^2}\,.
\end{align}
An appropriate ansatz for a WKB approximation to the solution is 
\begin{align}\label{eqn:WKB-ansatz}
 \hat h_\mathrm{WKB}(z)
 &=\exp \left\{ {\qq
     \left[
	 T_0(z)
	 +\qq^{-1}\,T_1(z)
	 +\qq^{-\alpha}\,T_\alpha(z)
	 +\cdots\right]} \right\} .
\end{align}
Subsequent terms in the exponent involve higher integer powers of $q^{-1}$
and $q^{-\alpha}$.
The ordering of the terms will be explained a-posteriori, when we find 
that $\alpha$ is non-integer and $1 < \alpha < 2$.
Inserting the expansion (\ref{eqn:WKB-ansatz}) into the radial equation
(\ref{eqn:radial-eq-for-WKB2}) and collecting like powers of $q$
produces the conditions:
\begin{align}\label{eqn:T-eq}
 (T_0')^2&=Q_0 \,,
 &
 T_0'' +2T_0' \, T_1' &=0 \,,
 &
 2T_0' \, T_\alpha' &= Q_\alpha \,.
\end{align}
Solving for $T_0$, $T_1$, and $T_\alpha$ yields two solutions
(due to the sign ambiguity in $\sqrt{Q_0}$).
One choice gives
\begin{align}
 T_0&=\int dz \> \sqrt{Q_0} \,,&
 T_1&=-\tfrac{1}{4}\log Q_0 \,,&
 T_\alpha&=\int dz \> \frac{Q_\alpha}{2\sqrt{Q_0}} \,,
\label{eq:T012}
\end{align}
where we define $\sqrt{Q_0(z)}$ as the branch which approaches
$
    +i s_0 / [2 \sqrt z \, (1{-}z^2)]
$
as $s_0 \to \infty$
(with $\sqrt z \ge 0$ for $z \in [0,1]$).
The other choice is obtained by replacing $\sqrt{Q_0}$ with $-\sqrt{Q_0}$.
The resulting
WKB approximations for two linearly independent solutions,
which we denote by $\hat h_{\rm WKB}^\pm$,
have the form
\begin{equation}\label{eqn:WKB-sol-leading}
    \hat h_{\rm WKB}^\pm(z)
    =
    Q_0(z)^{-1/4} \,
    e^{ \pm q \int^z dz' \>
	\left[
	    Q_0^{1/2} + \frac 12 q^{-\alpha} Q_\alpha Q_0^{-1/2} + \cdots \,
	\right]} \,.
\end{equation}
The most general solution is an arbitrary linear combination
of $\hat h_{\rm WKB}^\pm$.
Subleading terms in these WKB approximations are
negligible provided that
$|1-s_0^2-z^2|\, \gg \qq^{-\alpha}$
and $|1-s_0^2-z^2| \, z \gg \qq^{-2}$.
The first condition ensures that the $Q_0(z)$ term in
eq.~(\ref{eqn:radial-eq-for-WKB2}) is large compared to $Q_\alpha(z)$,
while the second condition ensures that $Q_0(z)$ also dominates the
$Q_2(z)$ term.

Near the horizon, $1{-}z \ll 1$,
we have $\sqrt{Q_0} \sim \frac{i}{4} s_0/(1{-}z)$
and $e^{ q \int^z dz' \> Q_0^{1/2} } \sim  (1{-}z)^{-\frac 14 i q s_0} $.
Hence
\begin{equation}
    \hat h_{\rm WKB}^+(z) \sim (1{-}z)^{\frac 12 - \frac 14 i q s_0} \,,\qquad
    \hat h_{\rm WKB}^-(z) \sim (1{-}z)^{\frac 12 + \frac 14 i q s_0} \,.
\label{eq:WKB-near-horizon}
\end{equation}
Only the behavior of $\hat h_{\rm WKB}^+$ 
matches the required near-horizon condition (\ref{eqn:hhat-bcB}),
so this is the solution of interest.

Near the boundary, $z \ll 1$, we have
$
    \sqrt{Q_0} \sim -\frac 12 \sqrt{(1{-}s_0^2)/z}
$,
with $\sqrt{1{-}s_0^2}$ defined to be positive just above
the branch cut running from $-1$ to 1.
Hence
$
    {\int^z dz' \> Q_0^{1/2}}
    \sim
    {-\sqrt{(1{-}s_0^2) z} }
$ 
and
\begin{equation}
    \hat h_{\rm WKB}^+(z)
    \sim
    [z/(1{-}s_0^2)]^{1/4} \,
    e^{-q \sqrt{(1{-}s_0^2) z} } \,.
\label{eq:WKB-overlap}
\end{equation}
The form (\ref{eq:WKB-overlap}) cannot, however, be directly compared with
the required boundary condition (\ref{eqn:hhat-bc}),
as the WKB approximation (\ref{eqn:WKB-sol-leading}) is not valid all
the way to $z = 0$;
as noted above, the WKB approximation is limited to
$z \gg q^{-2}/|1{-}s_0^2|$.
Therefore, we must match the WKB solution to a suitable
near-boundary approximation.%
\footnote
    {%
    If $s_0^2\notin [0,1]$, then the WKB approximation is valid for all
    $z\in[\epsilon,1]$ for any $\epsilon \gg q^{-2}/|1{-}s_0^2|$.
    But if $s_0^2$ is real and lies inside the interval $[0,1]$
    then there is a quadratic turning point at 
    $z_* = 1{-}s_0^2$.
    The WKB approximation (\ref{eqn:WKB-sol-leading}) is not accurate
    in a neighborhood of this turning point.
    Nevertheless, this does not invalidate the following argument
    matching WKB and near-boundary approximations, as one may deform
    the contour in $z$ along which one works from the real interval
    $[0,1]$ to a complex contour which runs from 0 to 1 but avoids
    the turning point at $z_*$.
    This contour deformation argument does not apply
    when $s_0^2 \to 1$, as the endpoints of the contour in $z$
    are necessarily fixed at 0 and 1.
    }

Provided $s^2 \ne 1$,
the Schr\"odinger equation (\ref{eqn:radial-eq-for-WKB})
for $\hat h$ simplifies 
near the boundary, $z \ll 1$, to
\begin{equation}
    \hat h''
    =
    \left[ \tfrac 14 \, q^2 \, (1{-}s^2) \, z^{-1} + \tfrac 34 z^{-2}\right]
    \hat h \,,
\end{equation}
with solutions given by regular or irregular modified Bessel functions,
\begin{align}
    \hat h^{\rm reg}(z)
    &= 
    \sqrt{z}\, I_2\big(q \sqrt{\left(1{-}s^2\right) z}\big) \,,
    &
    \hat h^{\rm irr}(z)
    &= 
    \sqrt{z}\, K_2\big(q \sqrt{\left(1{-}s^2\right) z}\big) \,.
\label{eq:near-boundary-Bessel}
\end{align}
These forms are valid for $z \ll 1$, regardless of the size of $q^2 z$,
up to relative corrections of order $z^2$.
In the overlap region $1 \gg z \gg q^{-2}/|1{-}s_0^2|$,
both WKB and near-boundary approximations are valid.
Within this region, the arguments of the Bessel functions in
the near-boundary approximations (\ref{eq:near-boundary-Bessel}) are large
and these solutions behave as%
\footnote
    {%
    These asymptotic forms, and the following argument, are valid provided
    $
	\sqrt{(1{-}s^2)z}
    $
    has positive real part.
    As the phase of $\sqrt{1{-}s^2}$ varies away from zero,
    it is convenient to perform the matching on the ray
    $
	\arg z = -\arg \sqrt{1{-}s^2}
    $,
    along which the arguments of the modified Bessel functions remain real.
    }
\begin{subequations}\label{eq:Bessel-asymp}%
\begin{align}
    \hat h^{\rm reg}(z)
    &\sim 
    (2\pi q)^{-1/2} \, (z/(1{-}s^2))^{1/4} \, e^{q \sqrt{(1{-}s^2)z}}  \,,
    \\
    \hat h^{\rm irr}(z)
    &\sim 
    (2q/\pi)^{-1/2} \, (z/(1{-}s^2))^{1/4} \, e^{-q \sqrt{(1{-}s^2)z}} \,.
\end{align}
\end{subequations}
Comparing these forms to the WKB behavior (\ref{eq:WKB-overlap}),
one sees that $\hat h_{\rm WKB}^+$ is proportional to 
the near-boundary solution $\hat h^{\rm irr}$, not to $\hat h^{\rm reg}$.
However, only the regular near-boundary solution $\hat h^{\rm reg}$
satisfies the boundary condition (\ref{eqn:hhat-bc}) requiring
$\O(z^{3/2})$ behavior as $z \to 0$.
The irregular solution $\hat h^{\rm irr}$ diverges as $\O(z^{-1/2})$
as $z \to 0$, violating the required regularity condition.
Consequently, the assumption that $s_0^2 \ne 1$ is inconsistent
with the boundary conditions (\ref{eqn:hhat-bcs});
solutions which satisfy the boundary condition at one end of our
interval in $z$ fail to satisfy the required boundary condition
at the other end.
Therefore, the only solutions which satisfy both boundary conditions
must have $s_0^2 = 1$, implying that quasinormal mode frequencies
approach $\pm q$ as $q \to \infty$.

\subsection{Subleading behavior}\label{sec:subleading-q}

Specializing (without loss of generality) to the case of $s_0 = +1$,
the integrals appearing in the WKB functions
(\ref{eq:T012}) may be explicitly evaluated and give:
\begin{align}
    T_0(z) &= -\tfrac i2 \left[\tan^{-1}(\sqrt z) - \tanh^{-1}(\sqrt z) \right],
    &
    T_\alpha(z) &= s_\alpha \big[ T_0(z) - i z^{-1/2} \,\big] .
\end{align}
Hence, the relevant WKB solution has the form
\begin{align}\label{eqn:WKB-sol}
    \hat h_{\rm WKB}^+(z) &= 
    e^{-i\pi/4} \, z^{-1/4} \sqrt{2(1{-}z^2)} \,
    \exp\Big\{
	q \left[
	    (1+s_\alpha \, q^{-\alpha}) \, T_0(z)
	    - i s_\alpha \, q^{-\alpha} \, z^{-1/2} 
	    + \cdots
	\right]
    \Big\} \,.
\end{align}
As discussed above, neglected higher order terms 
are negligible provided $z^2\qq^{\alpha}\gg 1$ and $z\,\qq^2\gg 1$.
Once again, this solution will need to be matched, within a suitable
overlap region, to an appropriate near-boundary solution.
For $z \ll 1$,
$T_0(z) \sim \tfrac i3 \, z^{3/2}$
and
(with no assumption on the size of $z$ compared to inverse powers of $q$),
the WKB solution (\ref{eqn:WKB-sol}) behaves as
\begin{equation}\label{eqn:WKB-matching}
    \hat h_{\rm WKB}^+(z) \sim
    \sqrt 2 \, e^{-i\pi/4} \, z^{-1/4} \,
    \exp\left[
	\tfrac i3 \, q \, z^{3/2} -i s_\alpha \, q^{1-\alpha} \, z^{-1/2}
	\right] .
\end{equation}
We now turn to the near-boundary region.
Non-uniformity between the small $z$ and $s_0^2 \to 1$ limits cause the
near-boundary behavior for $s_0^2 = 1$ to be qualitatively different from
the previously discussed $s_0^2 \ne 1$ case.
So we must redo the analysis starting from eq.~(\ref{eqn:radial-eq-for-WKB})
and specializing to $s_0 = 1$.
Assuming $z\ll 1$ and $\qq\gg 1$
(but making no assumptions about products of the form $z^a\qq$), 
the Schr\"odinger equation (\ref{eqn:radial-eq-for-WKB}) simplifies to
\begin{align}\label{eqn:hh-matching}
    \hat h'' &=
    \left[
	-\tfrac 14 \, q^2 \, z
	-\half \, s_\alpha \, q^{2-\alpha} \, z^{-1}
	+\tfrac{3}{4} \, z^{-2}
    \right] \hat h \,.
\end{align}
It is helpful to introduce a rescaled coordinate,
\begin{equation}\label{eqn:y-def}
    y \equiv z \, q^{2/3} \,,
\end{equation}
so that $\tilde h(y) \equiv \hat h(z(y))$ satisfies
\begin{align}\label{eqn:hh-matching2}
    \tilde h'' &=
    \left[
	-\tfrac 14 \, y
	-\half \, s_\alpha \, q^{4/3-\alpha} \, y^{-1}
	+\tfrac{3}{4} \, y^{-2}
    \right] \, \tilde h \,.
\end{align}
In terms of this rescaled coordinate,
the small-$z$ form (\ref{eqn:WKB-matching}) of the WKB solution becomes
\begin{align}\label{eqn:WKB-matching2}
    \hat h_{\rm WKB}(z(y))
    \sim
    y^{-1/4}
    \exp\left[
	\tfrac{i}{3} \, y^{3/2} -i s_\alpha \, q^{4/3-\alpha} \, y^{-1/2}
    \right] ,
\end{align}
and is valid for $y^2 \gg q^{4/3-\alpha}$.
Clearly, if\,%
\footnote
    {%
    If $\alpha > 4/3$, then all dependence on $s_\alpha$ in
    eqs.~(\ref{eqn:hh-matching2}) and (\ref{eqn:WKB-matching2})
    vanishes in the $q \to \infty$ limit,
    and the solution to eq.~(\ref{eqn:hh-matching2}) which matches
    onto the WKB solution for large $y$ fails to satisfy
    the $\O(y^{3/2})$ regularity condition at $y = 0$.
    This shows that the ratio $s = \omega/q$ must deviate from unity
    by terms at least as large as $\O(q^{-4/3})$.
    }
\begin{equation}\label{eq:alpha}
    \alpha = \tfrac {4}{3} \,,
\end{equation}
then we have a consistent description for asymptotically large $q$:
the WKB solution has a universal small-$z$ form,
$
    \hat h_{\rm WKB}(z(y))
    \sim
    y^{-1/4}
    \exp\left[
	\tfrac{i}{3} \, y^{3/2} -i s_\alpha^\infty \, \, y^{-1/2}
    \right]
$,
valid for $y \gg 1$,
which can smoothly match onto a solution $\tilde h(y)$
of the $q$-independent near-boundary equation,
\begin{align}\label{eqn:hh-matching3}
    \tilde h'' &=
    \left[
	-\tfrac 14 \, y
	-\half \, s_\alpha^\infty \, y^{-1}
	+\tfrac{3}{4} \, y^{-2}
    \right] \, \tilde h \,.
\end{align}
To determine allowed values for the constant $s_\alpha^\infty$,
one must find solutions to eq.~(\ref{eqn:hh-matching3}) which
are $\O(y^{3/2})$ as $y \to 0$ and,
up to an irrelevant overall constant, approach
$
    y^{-1/4}
    \exp\left[
	\tfrac{i}{3} \, y^{3/2} -i s_\alpha^\infty \, \, y^{-1/2}
    \right]
$ when $y \gg 1$.

Although it may seem most natural to work
on the ray with $\arg y = 0$
(corresponding to the original physical domain of $z \in [0,1]$)
when performing this matching,
this is not required.
For reasons which will momentarily become apparent,
it is more convenient to work along the rotated ray
$
    \arg y = \pi /3
$.
So we define
\begin{equation}
    y \equiv e^{i \pi/3} \, w \,,
\label{eq:rotate}
\end{equation}
with $w$ real and positive.
Then $\h(w) \equiv \tilde h(y(w))$ satisfies
\begin{equation}
    \h'' =
    \left[
	\tfrac 14 \, w
	-\lambda \, w^{-1}
	+\tfrac{3}{4} \, w^{-2}
    \right] \, \h \,,
\label{eq:rotateeq}
\end{equation}
where $\lambda \equiv \half s_\alpha^\infty \, e^{i \pi/3}$.
Boundary conditions become
$
    \h(w) \sim w^{-1/4} \exp\big[{-}\tfrac 13 \, w^{3/2} -2\lambda \, w^{-1/2}\big]
$
for $w \gg 1$, and $\h(w) = \O(w^{3/2})$ as $w \to 0$.
In other words, by rotating the contour, our desired solution now vanishes exponentially
for large argument.
Moreover, with these boundary conditions
eq.~(\ref{eq:rotateeq}) is a self-adjoint eigenvalue problem.
Specifically, $\lambda$ is an eigenvalue of the self-adjoint positive operator
$
    \sqrt w
    \left(
	-\partial_w^2 + \tfrac 14 w + \tfrac 34 w^{-2}
    \right)
    \sqrt w
$.
From the form of the effective potential appearing in this operator,
it is clear that it has a pure point spectrum.
So the eigenvalues $\{ \lambda_n \}$ must form a discrete
set of real, positive values.
Consequently, the subleading asymptotic coefficient $s_\alpha^\infty$
must have the form
\begin{equation}
    s_\alpha^\infty = c_n \, e^{-i \pi /3} \,,
\label{eq:pi/3}
\end{equation}
with a real, positive sequence of values $\{ c_1, c_2, \cdots \}$
equal to twice the eigenvalues $\{ \lambda_n \}$.

\begin{table}
\setlength{\tabcolsep}{7.5pt}
\begin{tabular}{c|c||c|c||c|c||c|c|c|c}
\hline
\hline
\multicolumn{10}{c}{helicity $\pm 2$ modes}\\
\hline
\hline
 $n$  & $c_n$ &
 $n$  & $c_n$ &
 $n$  & $c_n$ &
 $n$  & $c_n$ &
 $n$  & $c_n$ \\
\hline
1  & 4.464041100 &
6  & 33.29797173 &
11 & 71.39462943 &
16 & 115.5907121 &
21 & 164.5420243
\\
2  & 9.155136716 &
7  & 40.32733993 &
12 & 79.80148278 &
17 & 125.0308016 &
22 & 174.8285514
\\
3  & 14.48139869 &
8  & 47.67478411 &
13 & 88.43518883 &
18 & 134.6522859 &
23 & 185.2685346
\\
4  & 20.32785188 &
9  & 55.31510291 &
14 & 97.28444537 &
19 & 144.4485811 &
24 & 195.8575945
\\
5  & 26.61804258 &
10 & 63.22753437 &
15 & 106.3392576 &
20 & 154.4136692 &
25 & 206.5916511
\\
\hline
\hline
\end{tabular}
\caption{%
    Values of the asymptotic spectral deviation coefficients $\{ c_n \}$
    for the first 25 helicity $\pm2$ quasinormal frequencies,
    where $\omega_n/q = 1 + c_n \, e^{-i \pi/3} \, q^{-4/3} + \O(q^{-2})$.
    All digits shown are accurate.
    }
\label{tab:s1-values}
\end{table}

The Schr\"odinger equation (\ref{eq:rotateeq}) has an irregular singular
point at $w=\infty$ along with a regular singular point at $w = 0$.
An analytic solution does not appear to be possible,
but solving this equation numerically is relatively straightforward.
We describe our numerical techniques in appendix~\ref{app:numerics}
and present the resulting values for the first 25 spectral deviation
coefficients $\{ c_n \}$ in table~\ref{tab:s1-values}.

The values of $c_n$ rapidly increase with increasing mode number $n$.
For $n \gg 1$, one may use a further WKB approximation to find the large $n$
asymptotics of these coefficients.
When the eigenvalue $\lambda$ is large,
a simple WKB approximation for high order eigenfunctions is valid
in regions where the potential
$
    \tfrac 14 \, w - \lambda w^{-1} + \tfrac 34 \, w^{-2}
$
is sufficiently slowly varying.
One must appropriately match to a near-boundary approximation
(given by a Bessel function) for small $w$,
and also match across the linear turning point at
$w \approx 2\sqrt\lambda$.
Details of this exercise are presented in appendix \ref{app:WKB}.
One finds that solutions satisfying the required boundary conditions
exist when
$
    c_n = c_n^\infty \, (1 + \O(1/n^{2}))
$,
where
\begin{equation}
    c_n^\infty \equiv K \, (2n+1)^{4/3} \,,\qquad
\label{eqn:s-alpha-asympt}
\end{equation}
with
\begin{equation}
    K
    \equiv
    \half
    \left[
	\sqrt{\pi} \>
	{\Gamma \left(\tfrac{7}{4}\right)}\big/
	{\Gamma \left(\tfrac{5}{4}\right)}
    \right]^{4/3}
    \approx 1.092535 \,.
\label{eq:K}
\end{equation}

\begin{figure}
\center
  \includegraphics[width=0.40\linewidth]{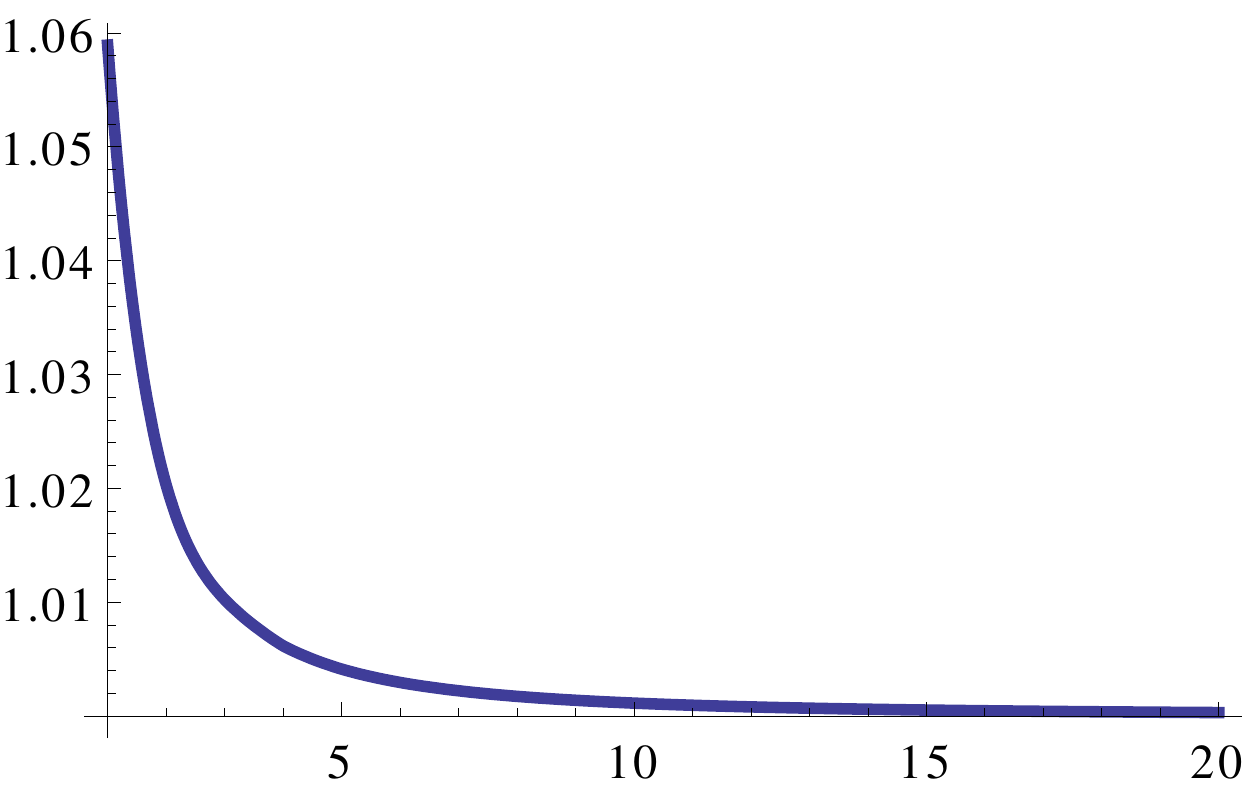} 
  \put(5,10){\small $n$}
  \put(-185,125){\small $c_n^\infty/c_n$}
\caption
    {%
    A plot of the ratio $c_n^\infty/c_n$
    of the asymptotic form (\ref{eqn:s-alpha-asympt}) to
    the numerical results in table~\ref{tab:s1-values}
    for the helicity $\pm2$ spectral deviation coefficients $c_n$.
    }
\label{fig:s-alpha-asympt}
\end{figure}

Figure~\ref{fig:s-alpha-asympt} shows a comparison of
this asymptotic form with the numerical
results in table~\ref{tab:s1-values}.
For the lowest $n = 1$ mode,
the deviation from the asymptotic scaling (\ref{eqn:s-alpha-asympt})
is approximately $6\%$
(far larger than the precision of the results in
table~\ref{tab:s1-values}).
But by $n = 5$ 
the asymptotic form is accurate to about half a percent.
The rapid approach to the asymptotic form (\ref{eqn:s-alpha-asympt})
could have been anticipated from the fact that already for
modest values of $n$ the coefficients $c_n$ become quite large
compared to unity.
Examination of the rate of convergence confirms the expected
$1/n^2$ scaling of the deviation.

To summarize, we have shown that helicity $\pm2$ quasinormal mode frequencies,
for large wavenumbers, have the form (\ref{eqn:s-exp}) with
$\alpha = 4/3$ and $s_\alpha^\infty$ having phase $-\pi/3$.
Continuing the WKB analysis, it is straightforward to show that the
next term in the large-$q$ expansion is $\O(q^{-1})$.
Therefore, for large wavenumbers, helicity $\pm2$
quasinormal mode frequencies are given by
\begin{equation}
    \omega_n(q) = q + c_n \, e^{-i \pi/3} \, q^{-1/3} + \O(q^{-1})\,,
\label{eqn:disp-asymp}
\end{equation}
plus reflected frequencies  $-\omega_n(q)^*$,
with the real coefficients $\{ c_n \}$ shown in table~\ref{tab:s1-values}.
These coefficients have the
large order asymptotic form (\ref{eqn:s-alpha-asympt}).
Restoring factors of $\pi T$ gives the result (\ref{eq:leadingform})
quoted in the introduction.

\section{Quasinormal mode frequencies: numerics}\label{sec:numerics}

To validate the large-$q$ asymptotics (\ref{eqn:disp-asymp}) and examine
the accuracy of this approximation for intermediate ranges of wavenumber,
we use pseudo-spectral methods \cite{Boyd:2001} to solve numerically the 
quasinormal mode equations for a wide range of wavenumbers.%
\footnote
    {%
    The fact that equation (\ref{eqn:radial-eq-for-WKB}) is of the Heun type
    can be used to derive an algebraic continued-fraction equation 
    satisfied by the quasinormal mode frequencies.
    We have used this to independently validate our numerical results 
    which were obtained by solving the
    differential equation (\ref{eqn:radial-eq}) using pseudo-spectral methods.
    However, the spectral approach proved to be computationally more robust.
    }
This extends previous work in ref.~\cite{Kovtun:2005ev}.
We consider first the helicity $\pm2$ case, and then examine helicity
$\pm1$ and 0.

\subsection{Helicity \texorpdfstring{$\pm2$}{+/-2}}\label{sec:numerics2}

As previously noted, frequencies for which the helicity $\pm2$ quasinormal
mode equation (\ref{eqn:radial-eq}) has solutions satisfying the required
regularity conditions at horizon and boundary come in pairs with opposite
real parts (and identical imaginary parts):
$\{ \omega_n \}$ and $\{ -\omega_n^* \}$.
So it is sufficient to consider only the positive frequency spectrum,
\ie, $\Re \omega \ge 0$.

To apply spectral methods, it is convenient to return to the original
form (\ref{eqn:radial-eq}) of the helicity $\pm2$ QNM equation.
Representing $h$ as a (truncated) series of Chebyshev polynomials,
\begin{equation}
    h(u) = \sum_{k=0}^M \> f_k \> T_k(2u{-}1) \,,
\label{eq:Chebexp}
\end{equation}
automatically satisfies the required regularity conditions at $u = 0$ and 1.
Demanding that equation (\ref{eqn:radial-eq}) [multiplied by $u(1{-}u^2)$]
be satisfied at each point $u = u_k$ on the collocation grid,
\begin{equation}
    u_k \equiv \half \left[1 - \cos (k \pi/M)\right] ,
\label{eq:grid}
\end{equation}
for $k = 0,{\cdots}, M$, yields a finite set of linear equations of the form
\begin{equation}
    (A - \omega B) \, \vec f = 0\,,
\label{eq:geneig}
\end{equation}
where,
given an explicit choice of the wavevector $\q$, $A$ and $B$ are numerical
$(M{+}1)\times (M{+}1)$ matrices.
The generalized eigenvalue problem (\ref{eq:geneig})
may be efficiently solved in $\O(M^3)$ time using standard methods.
Results for the first fifteen helicity $\pm2$
quasinormal mode frequencies $\omega_n(q)$
[or rather, the deviation ($\omega_n(q) {-} q$)],
for wavevectors $q = 10$, 20, 40, 80 and 160,
are shown in table~\ref{tab:frequencies2} of appendix~\ref{app:table}.

\begin{figure}
\center
  \includegraphics[width=0.40\linewidth]{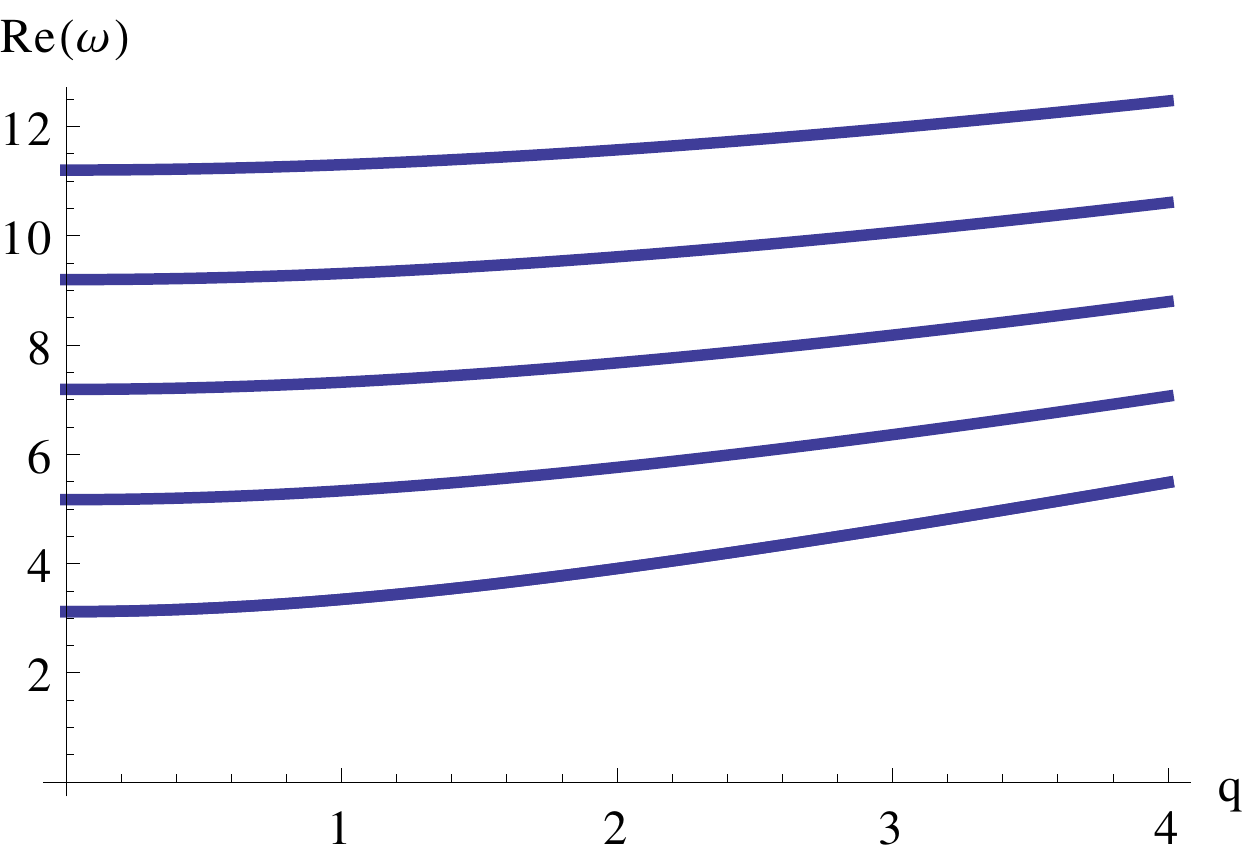} 
\qquad\quad
  \includegraphics[width=0.40\linewidth]{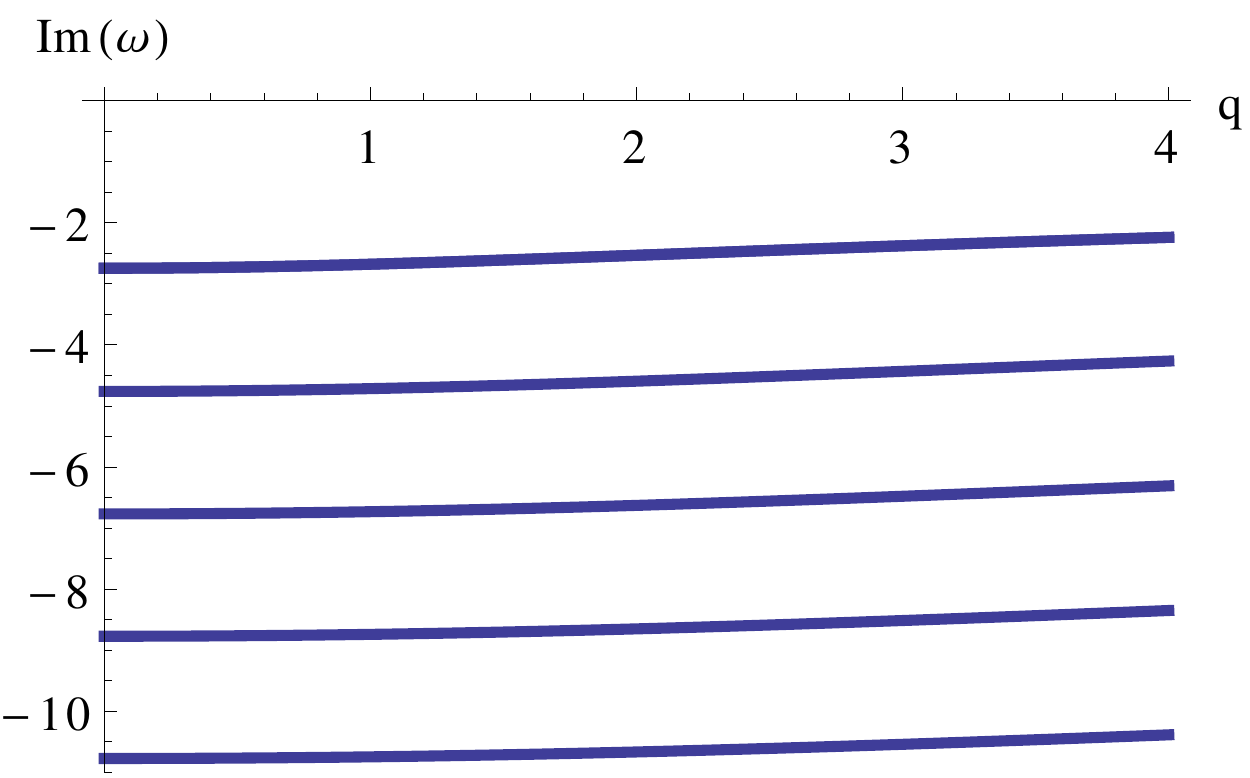} 
\caption
    {%
    Real (left) and imaginary (right) parts of the first 
    five helicity $\pm 2$ quasinormal frequencies,
    in units of $\pi T$,
    for small and intermediate wavevectors, $q \le 4\pi T$.
    }
\label{fig:spin-2-small-q}
\end{figure}

\begin{figure}
\center
  \includegraphics[width=0.45\linewidth]{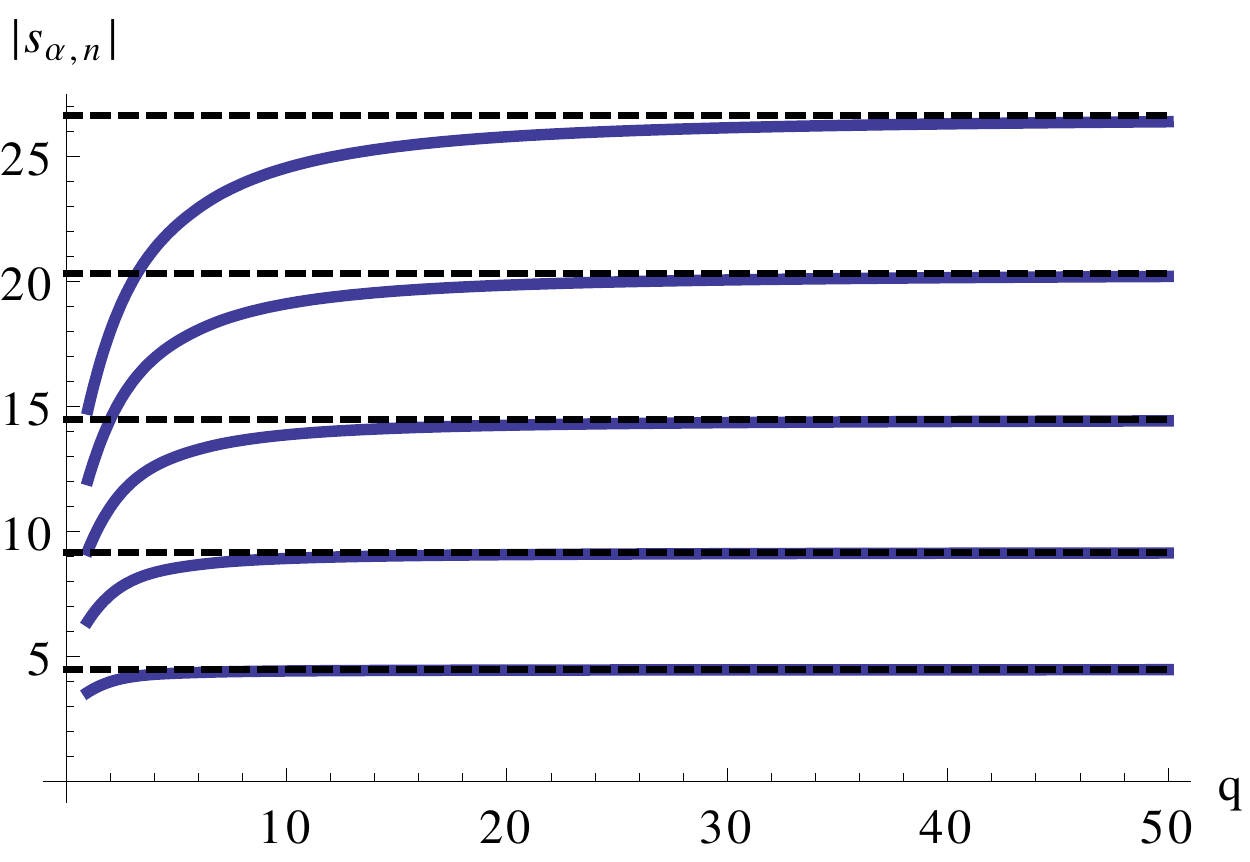} 
\qquad
  \includegraphics[width=0.45\linewidth]{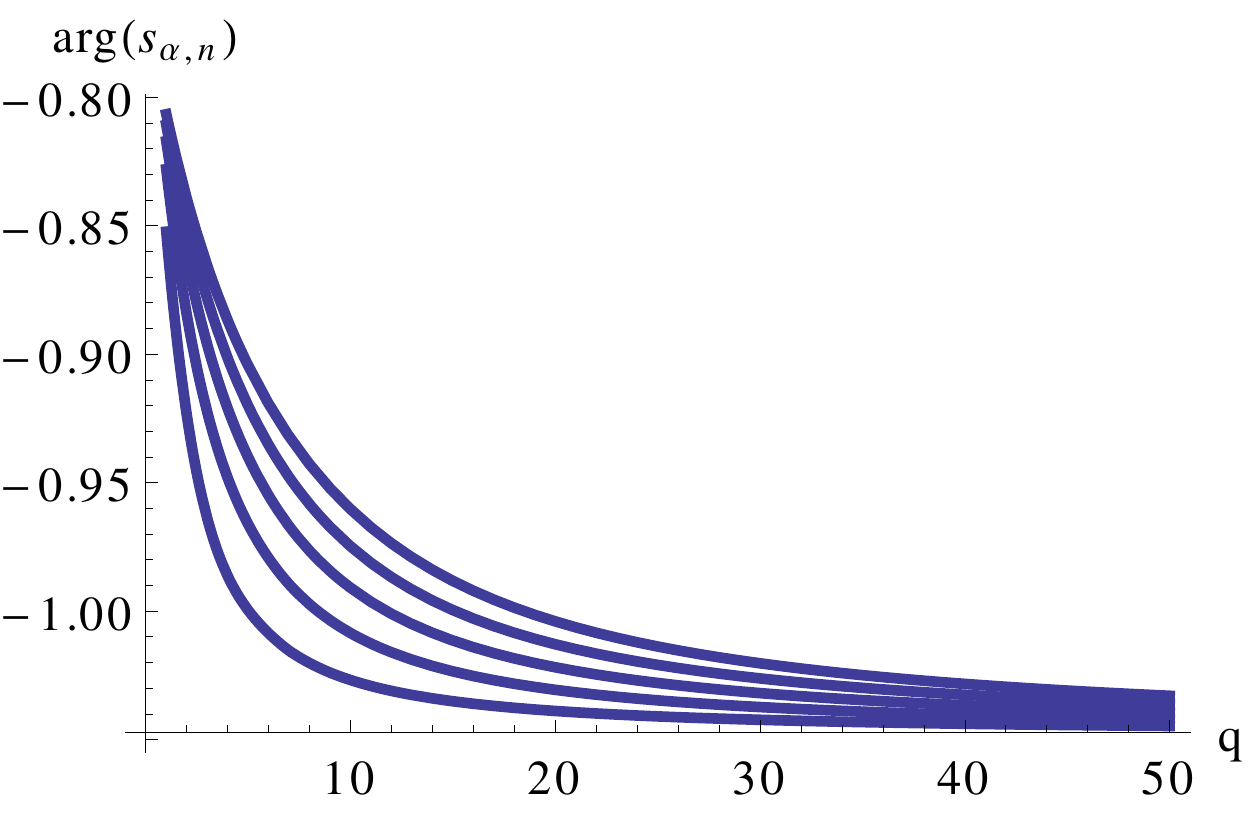} 
\caption
    {%
    Modulus (left) and phase (right) of the dispersive correction function
    $s_{\alpha,n}(q)$
    for the first five helicity $\pm2$ quasinormal modes,
    as a function of wavenumber $q$ (in units of $\pi T$).
    The complete quasinormal mode frequency is related
    to $s_{\alpha,n}(q)$ by eq.~(\ref{eqn:disp-ctilde}).
    The horizontal dashed lines show the asymptotic values $c_n$
    given in table~\ref{tab:s1-values}.
    For each mode,
    one sees that the magnitude of $s_{\alpha,n}(q)$ approaches
    the asymptotic value $c_n$
    while the phase approaches $-\pi/3$.
    Convergence is fastest for the lowest modes.
    }
\label{fig:intermediate-q}
\end{figure}

The real and imaginary parts of the first five helicity $\pm2$
quasinormal modes are plotted
in figure \ref{fig:spin-2-small-q}
for modest wavenumbers up to $4 \pi T$.
We have verified that our quasinormal frequencies
for $\qq{=}2$ agree with those given in app.~B of ref.~\cite{Kovtun:2005ev}.%
\footnote
    {%
    Note that Kovtun and Starinets \cite{Kovtun:2005ev} give results in
    units of $2\pi T$, not $\pi T$.
    }
To present results for larger wavenumbers in a manner which allows
easy comparison with the asymptotic form (\ref{eqn:disp-asymp}),
we use the definition (\ref{eqn:s-exp}) of the dispersive
correction function 
(with $\alpha = 4/3$),
repeated here,
\begin{equation}
    \omega_n(q) \equiv q + s_{\alpha,n}(q) \, q^{-1/3} \,.
\label{eqn:disp-ctilde}
\end{equation}
The magnitude and phase of the dispersive correction $s_{\alpha,n}(q)$ for the first 5 modes
are shown in figure~\ref{fig:intermediate-q}
for wavenumbers up to $q/(\pi T) = 50$.
One sees, as expected, that $s_{\alpha,n}(q)$ approaches the asymptotic value
$c_n \, e^{-i \pi/3}$ extracted from the WKB analysis.
Lower modes converge faster than higher modes.
The rapid rise of the magnitude $|s_{\alpha,n}(q)|$ as $q$ increases
from zero is an artifact of definition (\ref{eqn:disp-ctilde})
(since $\omega_n(q)$ has a finite $q \to 0$ limit).
But the leveling off of the magnitude after this rise provides a clear visual
indicator of the onset of the asymptotic regime.
From the figure it might appear that the convergence of the magnitude of
$s_{\alpha,n}(q)$ toward its asymptotic value $c_n$ is
considerably faster than the convergence of the phase to $-\pi/3$.
This, however, is an illusion produced by the rather compressed range
of the ordinate in the right hand plot
(which was chosen to make the different phase curves visually distinct).

One may parameterize the raw data in table \ref{tab:frequencies2} of appendix \ref{app:table}
using the functional form
\begin{equation}
    \omega_n(q) - q
	= A_n^{(1)} \, q^{-1/3}
	+ A_n^{(2)} \, q^{-1}
	+ A_n^{(3)} \, q^{-5/3}
	+ A_n^{(4)} \, q^{-7/3}
	+ A_n^{(5)} \, q^{-3}
    ,
\label{eq:parameterize}
\end{equation}
and demanding that the result reproduce the values in table~\ref{tab:frequencies2}.
This form is a truncation of the series which is generated by higher order asymptotic 
analysis.\footnote{%
  The powers of $q$ in the $A_n^{(1)}$ and $A_n^{(2)}$ terms reflect
  the result (\ref{eqn:disp-asymp}) of sec.~\ref{sec:subleading-q}.
  When recast as an expansion of $\omega(q)/q$, 
  higher order terms involve products of positive integer powers of
  $q^{-4/3}$ and $q^{-2}$
  arising from the decomposition (\ref{eqn:radial-eq-for-WKB2}) of the effective potential,
  and form a series in integer powers of $q^{-2/3}$.
}
The resulting values for the first coefficient $A_n^{(1)}$, when multiplied by $e^{i \pi/3}$,
provide independent estimates of the asymptotic coefficients $\{ c_n \}$.
These estimates, based on what is effectively an extrapolation to $q = \infty$, 
are less accurate than the values listed in table~\ref{tab:s1-values},
but the agreement is quite good.
The deviation is less than a part in $10^4$
for the first few modes, but grows to about half a percent for $n = 15$.
(This reflects the slower approach to the large-$q$ asymptotic form of progressively
higher modes.)
Moreover, we have explicitly tested that using the parameterization
(\ref{eq:parameterize})
of the data in table \ref{tab:frequencies2}, the resulting functions
reproduce the directly calculated values of the quasinormal mode frequencies
used to produce figure \ref{fig:intermediate-q}
(showing the range $10 \le q \le 50$)
to within a precision of two parts in $10^4$.

\subsection{Helicity \texorpdfstring{$\pm 1$}{+/-1} and 0}

To analyze perturbations with helicity $\pm 1$ and 0, it is convenient to use
the gauge invariant linear combinations of metric perturbations introduced
by Kovtun and Starinets \cite{Kovtun:2005ev}.
With a Fefferman-Graham form for the metric of the black brane geometry,
\begin{equation}
    ds^2 = \frac 1z \left[ -(1{-}z^2) \, d\tau^2 + d\x^2 \, \right]
    + \frac {dz^2}{4 z^{2} (1{-}z^2)} \,,
\label{eq:ds2FG}
\end{equation}
helicity $\pm1$ and 0 linear combinations are, respectively,
\begin{subequations}\label{eq:Z12}%
\begin{align}
    Z_1 &\equiv z \left( q \, \delta g_{\tau x_1}+\omega \, \delta g_{ux_1} \right),
\label{eq:Z1}
\\
    Z_2 &\equiv 
    z \left\{
	\omega^2 \, \delta g_{x_3 x_3}
	+ 2 \omega q \, \delta g_{\tau x_3}
	+ q^2 \, \delta g_{\tau\tau} +
	q^2 \left[(1{-}z^2) + 2 u^2 - \omega^2/q^2 \right]
	    \left(\delta g_{x_1x_1}+\delta g_{x_2x_2}\right)
    \right\} .
\label{eq:Z2}
\end{align}
\end{subequations}
Decoupled second order linear equations satisfied by these fluctuations were
derived in ref.~\cite{Kovtun:2005ev}.
Converting to our preferred infalling coordinates leads to the
following equations for these perturbations,
\begin{align}\label{eqn:radial-spin1}
    0 &=
    \widetilde Z_1''
    +
    \left[\frac{5}{u}+\frac{2i\ww}{f}-\frac{4u^3\ww^2}{f\,(\ww^2-q^2f)}\right]
    \widetilde Z_1'
    +
    \left[
	\frac{i\ww - u \qq^2}{u f}
	-4\ww\frac{4u^3\ww+i(\qq^2-\ww^2)}{u f \, (\ww^2-\qq^2 f)}
    \right] \widetilde Z_1 \,,
\\[5pt]
     0 &=
     \widetilde Z_2''
    +
    \left[
	\frac{1}{u}
	+\frac{2i\ww}{f}
	+\frac{4u^4(2\qq^2-3\ww^2)-12(\qq^2{-}\ww^2)f}
	    {u f \, (3\ww^2-(f{+}2)\,q^2)}
    \right]
    \widetilde Z_2'
\nonumber\\
    &\;\;\qquad {}
    +\left[
	-\frac{\qq^2}{f}
	+\frac{16(\qq^2-3\ww^2)-15i\ww \, (\qq^2{-}\ww^2)-3i\qq^2\ww u^4}
	    {u f \, (3\ww^2-(f{+}2)\,q^2)}
    \right]\widetilde Z_2 \,,
    \label{eqn:radial-spin0}
\end{align}
with $f(u) \equiv 1-u^4$.
Details of the transformation yielding these equations are given in
appendix \ref{app:transform}.
The required boundary conditions for
the functions $\widetilde Z_1(u)$ and $\widetilde Z_2(u)$ 
are just regularity at both horizon ($u{=}1$) and boundary ($u{=}0$).
Frequencies for which solutions satisfying these boundary conditions
exist are either pure imaginary, or else
come in pairs with opposite real parts, $\omega$ and $-\omega^*$.
Therefore, without loss of generality, in the following discussion
we consider $\Re \omega \ge 0$.

After multiplying the helicity $\pm 1$ equation (\ref{eqn:radial-spin1})
by its frequency-dependent denominator $\omega^2 - q^2 f$,
and likewise multiplying the helicity 0 equation (\ref{eqn:radial-spin0})
by $3 \omega^2 - (f{+}2) \, q^2$, both equations become
cubic generalized eigenvalue problems of the form
\begin{equation}\label{eqn:generalized-cubic}
    \left(
	\omega^3 \, O_3
	+\omega^2 \, O_2
	+\omega \, O_1
	+ O_0
    \right) \widetilde Z
    =0 \,,
\end{equation}
where each $O_i$ is a linear operator.
By replicating the function space on which one works,
this may be converted into a conventional generalized eigenvalue problem,
$A \, X = \omega \, B \, X$,
where
$X \equiv (\omega^2 \widetilde Z,\omega \widetilde Z, \widetilde Z)$ and%
\footnote
    {%
    This procedure is just a restatement of the fact that a single
    linear equation third order in time derivatives can be
    converted into a system of three coupled equations,
    each first order in time derivatives.
    }
\begin{equation}\label{eqn:generalized-cubic-to-linear}
    A \equiv
    \begin{pmatrix}
	O_2 & O_1 & O_0\\
	-\mathds{1} & 0 & 0\\
	0 & -\mathds{1} & 0 
    \end{pmatrix}
\,,\qquad
    B \equiv
    \begin{pmatrix}
	-O_3 & ~0~ & ~0~ \\
	0 & \mathds{1} & 0\\
	0 & 0 & \mathds{1}
    \end{pmatrix}
    \,.
\end{equation}
Applying pseudo-spectral methods to convert the linear radial differential
operators $O_i$ into matrices and solving the resulting finite dimensional 
generalized eigenvalue problem proceeds in the same manner described previously.
Results for the first fifteen helicity $\pm1$ and 0
quasinormal mode frequencies $\omega_n(q)$
[or rather, the deviation ($\omega_n(q) {-} q$)],
for wavevectors $q = 10$, 20, 40, 80 and 160,
are given in tables \ref{tab:frequencies1} and \ref{tab:frequencies0}
of appendix~\ref{app:table}.

\begin{figure}
\center
  \includegraphics[width=0.40\linewidth]{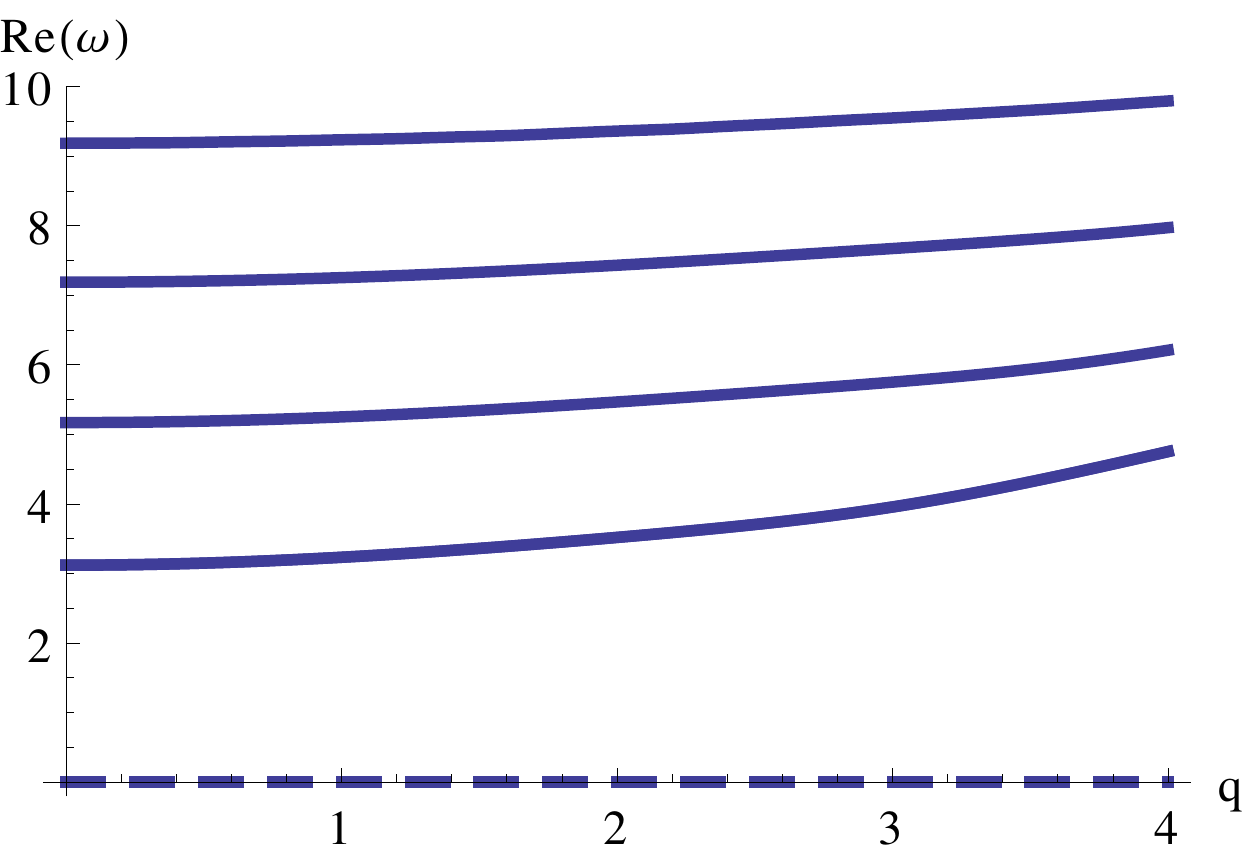} 
\qquad\quad
  \includegraphics[width=0.40\linewidth]{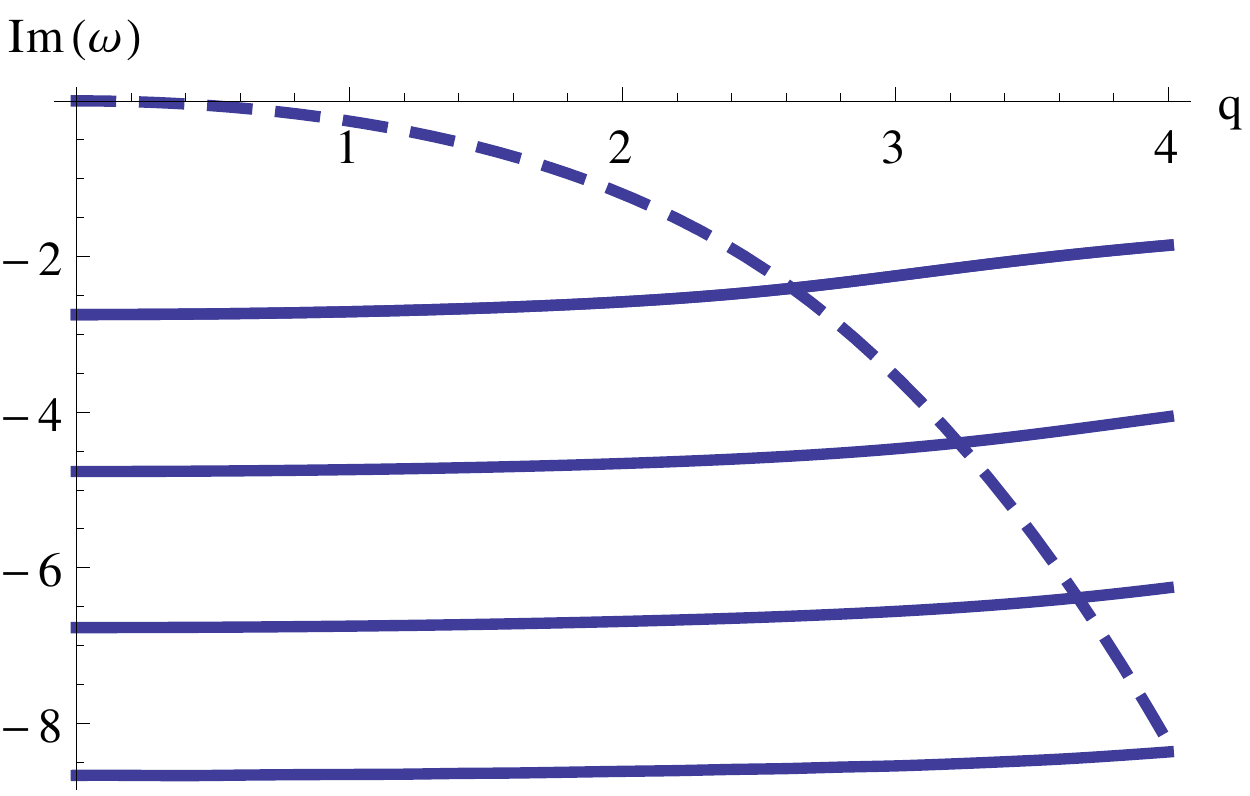} 
\caption
    {%
    Real (left) and imaginary (right) parts of the first five
    helicity $\pm 1$ quasinormal mode frequencies
    in units of $\pi T$,
    for small and intermediate wavevectors, $q \le 4\pi T$.
    There is one diffusive mode with pure imaginary frequency
    which approaches zero as $q \to 0$.
    The frequency of this hydrodynamic shear mode is shown with
    a dashed curve in both plots.
    }
\label{fig:spin-1-small-q}
\end{figure}

We first discuss helicity $\pm1$ perturbations.
The real and imaginary parts of
the first five quasinormal frequencies are plotted
in fig.~\ref{fig:spin-1-small-q} for $q \le 4\pi T$.%
\footnote
    {%
    Our numerical results are consistent with
    the values given for the non-hydrodynamic quasinormal modes in app.~B
    of ref.~\cite{Kovtun:2005ev}.
    (Hydrodynamic modes were excluded from their table.)
    For the hydrodynamic modes at $q = 2\pi T$,
    we find $\omega/(\pi T)=-1.19613i$ 
    for the diffusive purely imaginary helicity $\pm 1$ shear mode,
    and 
    $\omega/(\pi T)=1.48286 - 0.57256 i$ 
    for the helicity 0 propagating sound mode.
    }
For modest wavenumbers, $q \lesssim 2.6 \, \pi T$,
the most weakly damped mode
is a hydrodynamic shear mode whose frequency is pure imaginary
and vanishes as $q \to 0$.
This frequency, which is shown by dashed lines in
fig.~\ref{fig:spin-1-small-q},
moves down the imaginary axis as $q$ increases.
As seen in the figure and noted in ref.~\cite{Amado:2008ji},
the frequency of this mode crosses the imaginary 
parts of other mode frequencies (having non-zero real parts)
at various intermediate values of $q$.
For $q \gg T$, this mode becomes highly damped and is not
among the minimally damped modes discussed below.

\begin{figure}
\center
  \includegraphics[width=0.40\linewidth]{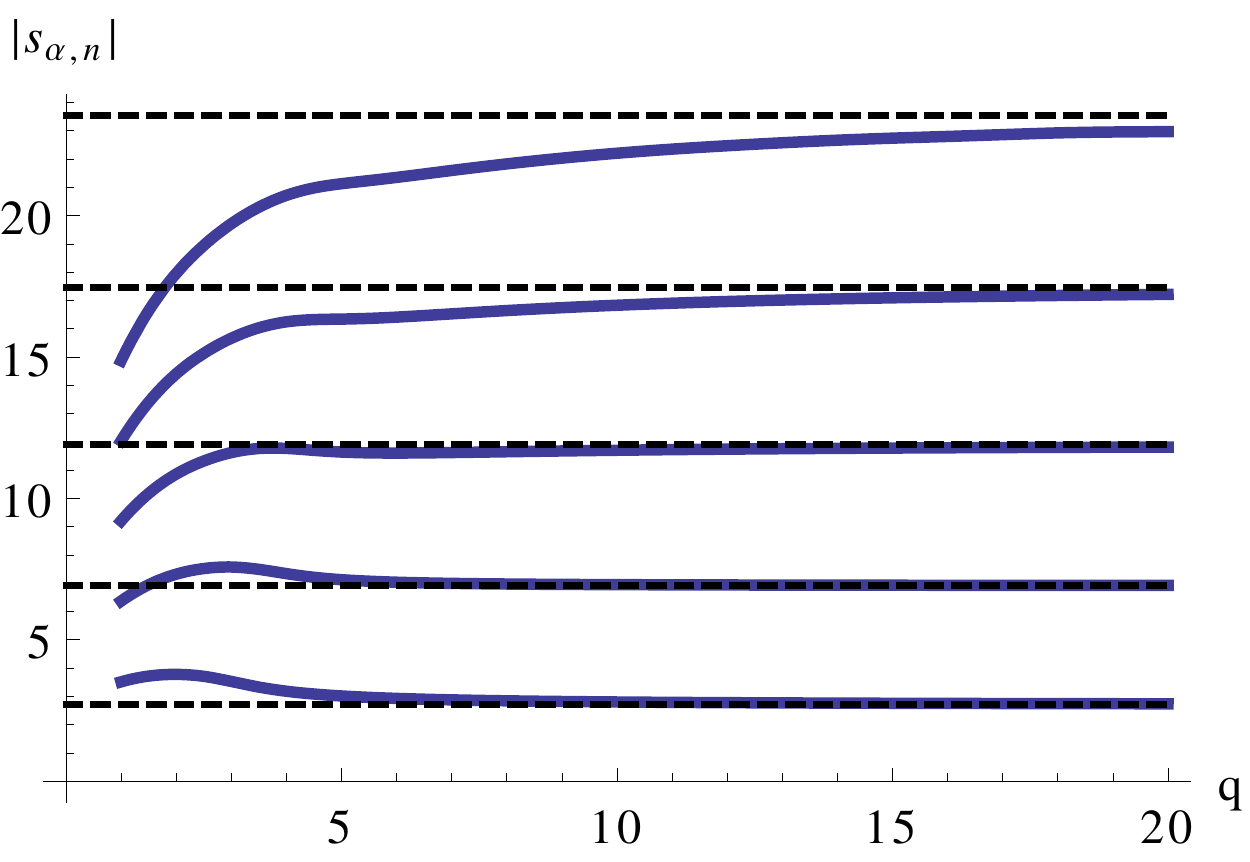} 
\qquad\quad
  \includegraphics[width=0.40\linewidth]{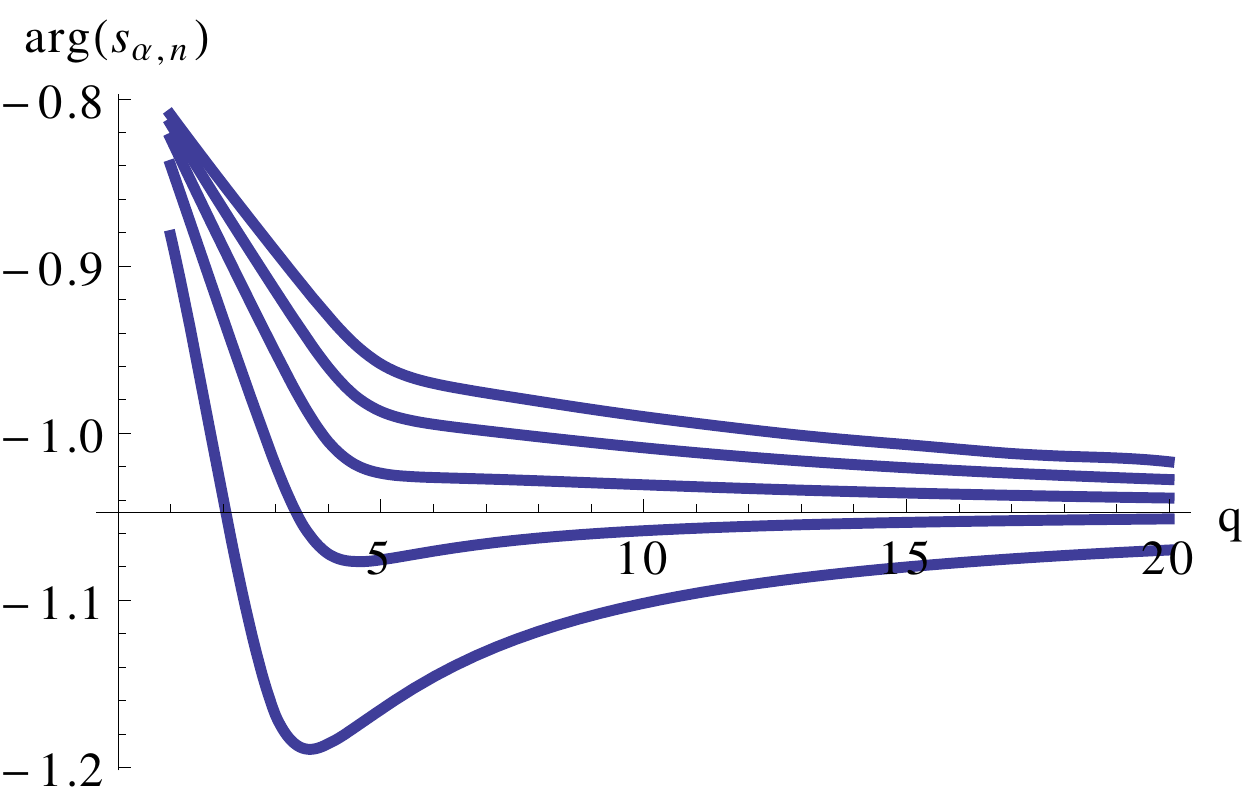} 
\caption
    {%
    Modulus (left) and phase (right) of the dispersive correction 
    $s_{\alpha,n}(q)$ for the first five (non-hydrodynamic) helicity $\pm 1$ quasinormal modes,
    as a function of wavenumber $q$ (in units of $\pi T$).
    The complete quasinormal mode frequency is related
    to $s_{\alpha,n}(q)$ by eq.~(\ref{eqn:disp-ctilde}).
    For each mode, one sees that the magnitude $|s_{\alpha,n}(q)|$
    becomes approximately constant as $q$ increases, and the corresponding
    phase approaches a value close to $-\pi/3$.
    Horizontal dashed lines show the asymptotic values extracted using
    the parameterization (\ref{eq:parameterize}) applied to the data in
    table \ref{tab:frequencies1}, and listed in table \ref{tab:s1-values-spin1}.
    Near-asymptotic behavior sets in for moderate values of wavevector,
    $q/(\pi T) \approx 5$.
    }
\label{fig:spin-1-large-q}
\end{figure}

To examine larger values of $q$ and the approach to the asymptotic regime,
we plot in fig.~\ref{fig:spin-1-large-q}
the magnitude and phase of the dispersive correction $s_{\alpha,n}(q)$,
defined via eq.~(\ref{eqn:disp-ctilde}),
of the lowest five helicity $\pm 1$ modes
(excluding the hydrodynamic shear mode)
for $q/(\pi T)$ up to 20.
Unlike the helicity $\pm 2$ case, 
one sees non-monotonic behavior in the lowest modes as $q$ increases.
Although we have not done an independent WKB calculation for helicity $\pm 1$
to determine asymptotic values directly,
from the plots it certainly appears that the magnitudes $|s_{\alpha,n}(q)|$
are approaching constant values while all phases are converging to a value
near $-\pi/3$.
Near-asymptotic behavior begins to be apparent for quite modest values of
wavenumber, $q/(\pi T) \approx 5$.

\begin{table}
\setlength{\tabcolsep}{10pt}
\begin{tabular}{c|c|c||c|c|c||c|c|c}
\hline
\hline
\multicolumn{9}{c}{helicity $\pm 1$ modes}\\
\hline
\hline
$n$ & $|c_n|$ & $\arg(c_n)$ &
$n$ & $|c_n|$ & $\arg(c_n)$ &
$n$ & $|c_n|$ & $\arg(c_n)$
\\
\hline
1& $2.69717$ & $0.000001$ & 6  & $30.0107$ & $0.00051$ & 11 & $67.2833$ & $0.00283$\\
2& $6.90578$ & $0.000003$ & 7  & $36.8644$ & $0.00087$ & 12 & $75.5084$ & $0.00325$\\
3& $11.8887$ & $0.000022$ & 8  & $44.0488$ & $0.00133$ & 13 & $83.9421$ & $0.00354$\\
4& $17.4637$ & $0.000093$ & 9  & $51.5313$ & $0.00183$ & 14 & $92.5695$ & $0.00369$\\
5& $23.5271$ & $0.000248$ & 10 & $59.2841$ & $0.00235$ & 15 & $101.379$ & $0.00366$\\
\hline
\hline
\end{tabular}
\caption
    {%
    Estimates for the magnitude and phase of the asymptotic
    spectral deviation coefficients $\{ c_n \}$ for the first fifteen
    helicity $\pm 1$ quasinormal mode frequencies,
    extracted from the parameterization (\ref{eq:parameterize})
    of the helicity $\pm1$ data in table \ref{tab:frequencies1}
    of appendix \ref{app:table}.
    Within the accuracy of the parameterization,
    the phases of $c_n$ are all compatible with zero.
    }
\label{tab:s1-values-spin1}
\end{table}

One may parameterize the helicity $\pm1$ data in table \ref{tab:frequencies1} of
appendix \ref{app:table} using the same functional form (\ref{eq:parameterize})
suggested by the helicity $\pm2$ WKB analysis.
The resulting parameterizations reproduce
the directly calculated values of quasinormal mode frequencies
used to produce figure \ref{fig:spin-1-large-q}
(showing the range $10 \le q \le 20$)
to within a precision of five parts in $10^4$.
Although not a formal proof,
the consistency and accuracy of the parameterization (\ref{eq:parameterize}),
when applied to our helicity $\pm1$ data,
strongly suggests that helicity $\pm 1$ quasinormal mode frequencies have
the same large-$q$ asymptotic form (\ref{eqn:disp-asymp}) as do helicity $\pm 2$ modes.
The first coefficients $\{ A_n^{(1)} \}$ of the parameterization, when multiplied
by $e^{i\pi/3}$,
directly give estimates of the asymptotic spectral deviation coefficients $\{ c_n \}$ for helicity
$\pm1$ modes.
Table \ref{tab:s1-values-spin1} lists these estimates for the first fifteen modes.
Within the accuracy of the parameterization
(as determined by our tests with $10 < q < 20$),
the phases of the asymptotic coefficients $c_n$ are all compatible with zero.

\begin{figure}
\center
  \includegraphics[width=0.40\linewidth]{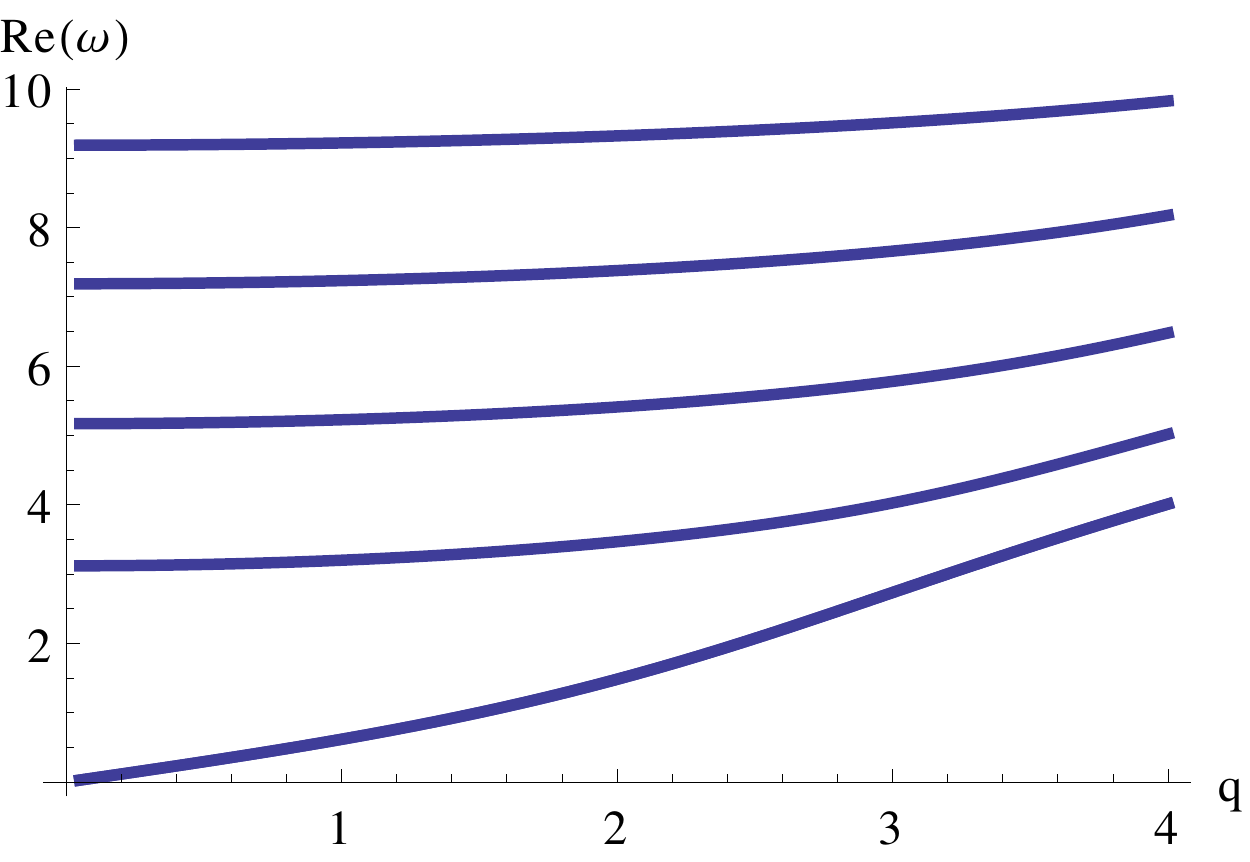} 
\qquad\quad
  \includegraphics[width=0.40\linewidth]{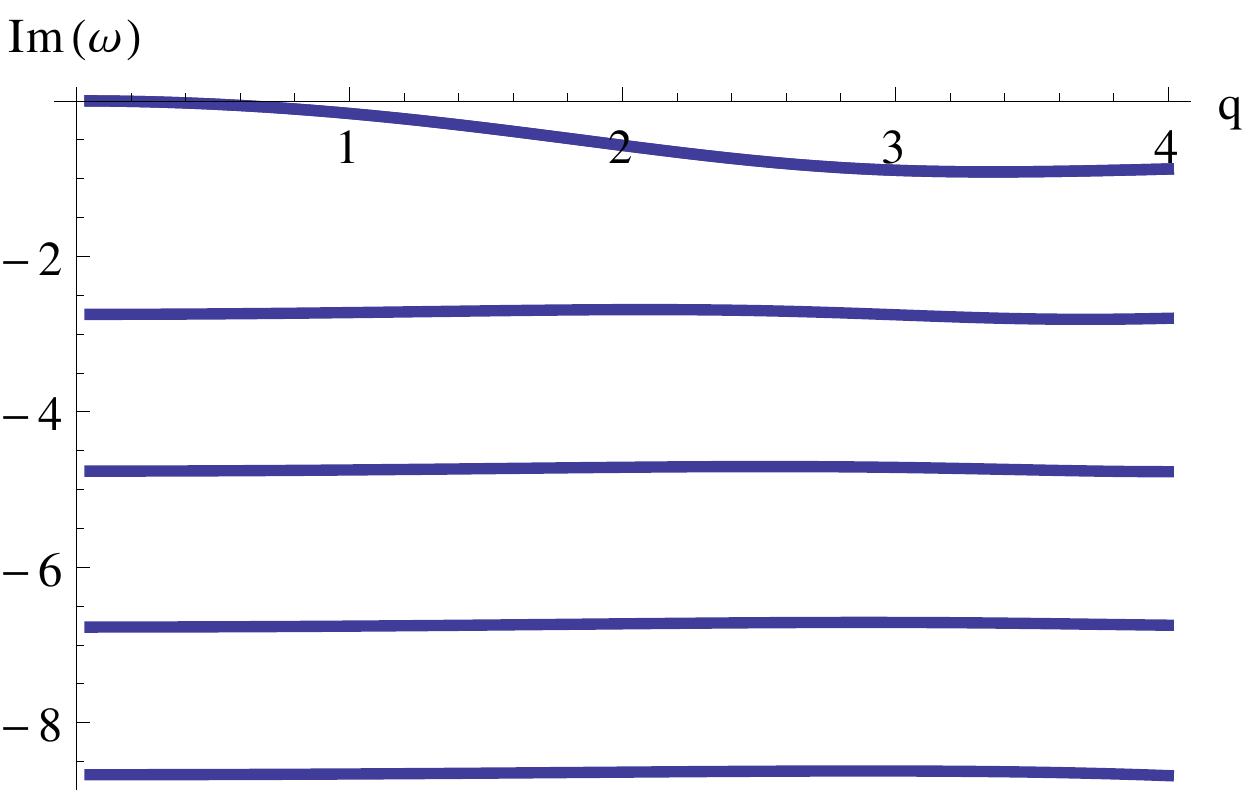} 
\vspace*{-7pt}
\caption
    {%
    Real (left) and imaginary (right) parts of the
    first five helicity 0 quasinormal frequencies
    in units of $\pi T$, for
    small and intermediate wavevectors, $q \le 4 \pi T$.
    There is one hydrodynamic (sound) mode whose frequency
    vanishes as $q \to 0$.
    }
\label{fig:spin-0-small-q}
\end{figure}

\begin{figure}
\center
  \includegraphics[width=0.40\linewidth]{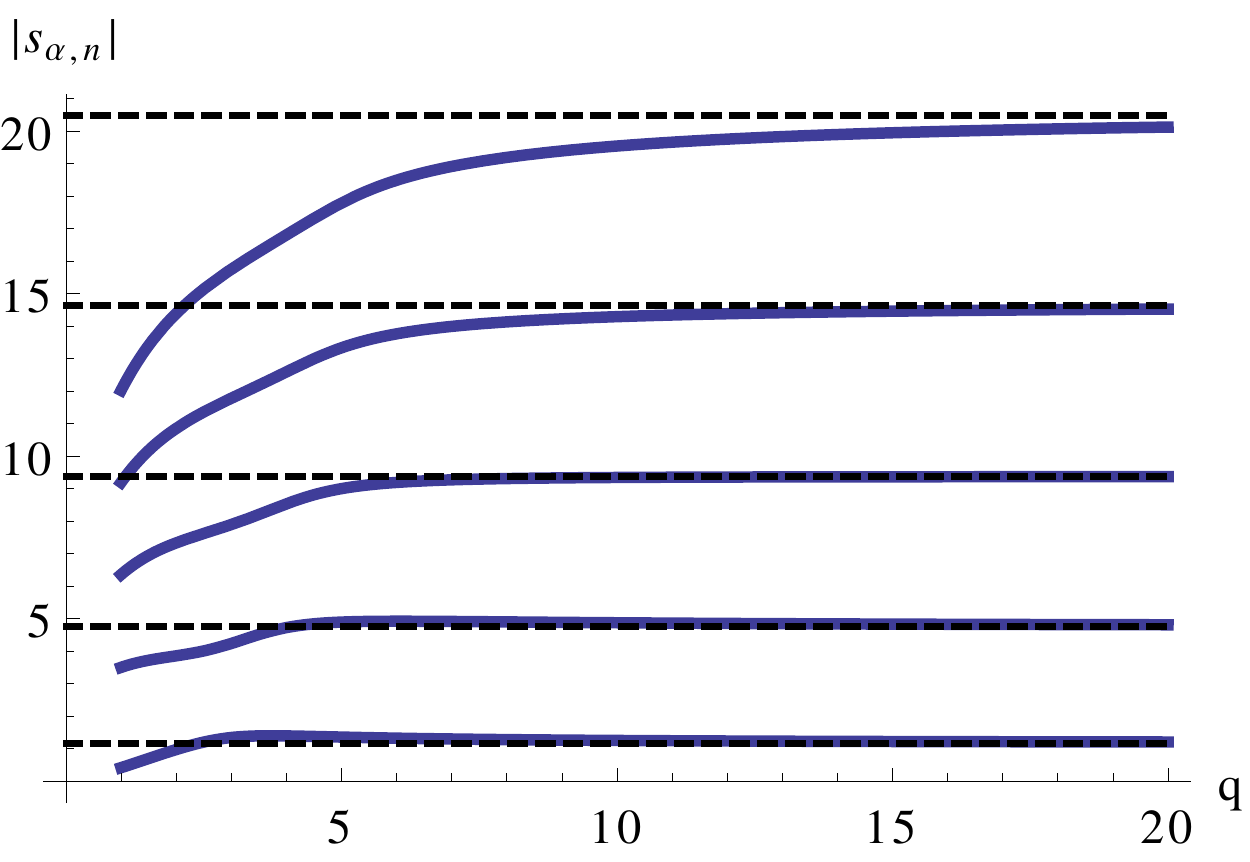} 
\qquad\quad
  \includegraphics[width=0.40\linewidth]{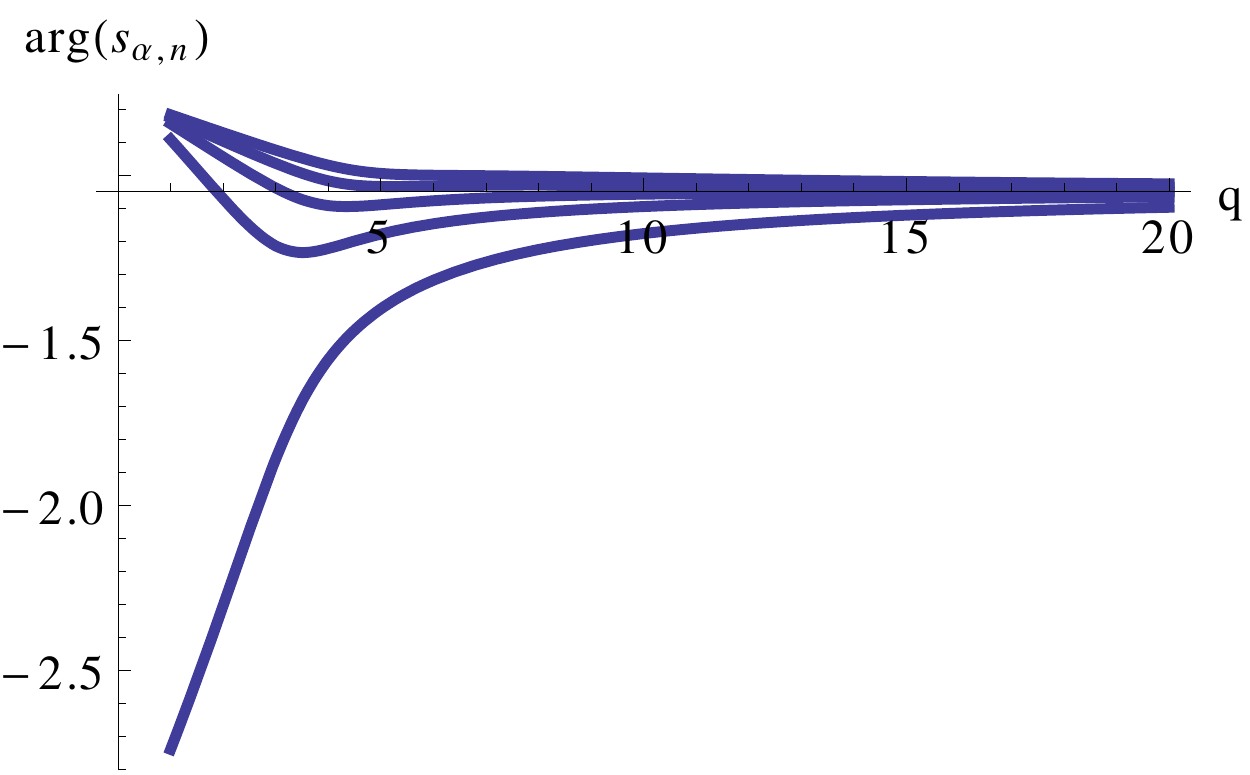} 
\caption
    {%
    Modulus (left) and phase (right) of the dispersive correction 
    $s_{\alpha,n}(q)$
    for the first five helicity 0 quasinormal modes,
    as a function of wavevector $q$ (in units of $\pi T$).
    The complete quasinormal mode frequency is related to $s_{\alpha,n}(q)$
    by eq.~(\ref{eqn:disp-ctilde}).
    Once again, for each mode one sees that the magnitude $|s_{\alpha,n}(q)|$
    becomes approximately constant as $q$ increases, and the corresponding
    phase approaches a value close to $-\pi/3$.
    Horizontal dashed lines show the asymptotic values extracted using the
    parameterization (\ref{eq:parameterize}) applied to the data in table
    \ref{tab:frequencies0}, and listed in table \ref{tab:s1-values-spin0}.
    Near-asymptotic behavior sets in for $q/(\pi T) \approx 5$.
    }
\label{fig:spin-0-large-q}
\end{figure}

We now turn to helicity 0 modes, whose behavior largely parallels that of the
helicity $\pm 1$ modes just discussed.
Fig.~\ref{fig:spin-0-small-q} plots the real and imaginary parts
of the first five helicity 0 quasinormal modes for $q/(\pi T) \le 4$.
There is one hydrodynamic helicity 0 (sound) mode, whose frequency
vanishes as $q \to 0$
(with $\Re \omega = \O(q)$ and $\Im \omega = \O(q^2)$).
As noted in ref.~\cite{Chesler:2011nc}, the helicity 0
hydrodynamic sound mode smoothly evolves from small to large values of $q$
and always remains the most weakly damped mode.
Its damping, as measured by $\Im \omega / \Re \omega$,
rises linearly from $q = 0$, reaches a maximum at $q/(\pi T)\approx 2.120$,
and then decreases as $\O(q^{-4/3})$ as $q$ continues to grow.
Fig.~\ref{fig:spin-0-large-q} plots the modulus and phase of the spectral deviation
function $s_{\alpha,n}(q)$ for these modes out to $q/(\pi T) = 20$.
From the figure one sees, once again, that
the magnitudes $|s_{\alpha,n}(q)|$ are
nearly constant for $q/(\pi T) \gtrsim 5$ and all phases approach a value
close to $-\pi/3$.

As before, one may parameterize the helicity 0 data in table \ref{tab:frequencies0}
of appendix \ref{app:table} with the functional form (\ref{eq:parameterize}) used earlier. 
The resulting parameterizations reproduce directly calculated values of 
helicity 0 quasinormal mode frequencies for wavevectors throughout the range $10 \le q \le 20$
to within a precision of a part in $10^4$.
This consistency strongly suggests that helicity 0 quasinormal mode frequencies
also have the same asymptotic form (\ref{eqn:disp-asymp}).
Table \ref{tab:s1-values-spin0} shows our resulting estimates, extracted
from this simple parameterization, for the spectral deviation coefficients $\{ c_n \}$
for the first fifteen helicity 0 modes.
Within the accuracy of the parameterization, the phases of the
asymptotic coefficients $c_n$ are, once again, all compatible with zero.

\begin{table}
\setlength{\tabcolsep}{10pt}
\begin{tabular}{c|c|c||c|c|c||c|c|c}
\hline
\hline
\multicolumn{9}{c}{helicity $0$ modes}\\
\hline
\hline
$n$ & $|c_n|$ & $\arg(c_n)$ &
$n$ & $|c_n|$ & $\arg(c_n)$ &
$n$ & $|c_n|$ & $\arg(c_n)$
\\
\hline
1& $1.17236$ & $0.000003$ & 6  & $26.7489$ & $0.000363$ & 11 & $63.2685$ & $0.002594$\\
2& $4.76469$ & $0.000003$ & 7  & $33.4190$ & $0.000675$ & 12 & $71.3809$ & $0.003046$\\
3& $9.36737$ & $0.000010$ & 8  & $40.4393$ & $0.001086$ & 13 & $79.7103$ & $0.003402$\\
4& $14.6512$ & $0.000050$ & 9  & $47.7737$ & $0.001570$ & 14 & $88.2406$ & $0.003624$\\
5& $20.4734$ & $0.000158$ & 10 & $55.3920$ & $0.002087$ & 15 & $96.9584$ & $0.003684$ \\
\hline
\hline
\end{tabular}
\caption
    {%
    Estimates for the magnitude and phase of the asymptotic
    spectral deviation coefficients $\{ c_n \}$ for the first fifteen
    helicity $0$ quasinormal mode frequencies,
    extracted from the parameterization (\ref{eq:parameterize})
    of the helicity $0$ data in table \ref{tab:frequencies0}
    of appendix \ref{app:table}.
    Within the accuracy of the parameterization,
    the phases of $c_n$ are all compatible with zero.
    }
\label{tab:s1-values-spin0}
\end{table}

In summary, we have compelling evidence that,
for all helicities of metric perturbations,
quasinormal mode frequencies
have the large $q$ asymptotic form 
$
    \omega_n(q) = q + c_n \, e^{-i \pi/3} \, q^{-1/3} + \O(q^{-1})
$,
with $\O(q^{-4/3})$ relative corrections to a massless $\omega = \pm q$
dispersion relation, and with real positive coefficients $\{ c_n \}$
characterizing the dispersive correction.
This large-$q$ asymptotic form is reasonably accurate,
at least for low order modes, down to $q/(\pi T) \approx 5$.

\subsection{Large order asymptotics}

Comparing the helicity $\pm2$ spectral deviation coefficients listed in table \ref{tab:s1-values}
with our corresponding helicity $\pm1$ or 0 values shown in tables \ref{tab:s1-values-spin1}
and \ref{tab:s1-values-spin0},
it is clear by inspection that the helicity $\pm1$ and 0 spectral deviation coefficients grow
about as fast with increasing mode number as do the helicity $\pm2$ coefficients.
Given the known asymptotic behavior (\ref{eqn:s-alpha-asympt}) of the helicity $\pm2$
coefficients, it is natural to try fitting our estimates of helicity $\pm1$ and 0 spectral
deviation coefficients using a function of the form
$
    b \, (2n + a)^{4/3}
$.
It is also instructive, for comparison, to apply the same procedure to estimates of
the helicity $\pm2$ spectral deviation coefficients generated by the parameterization
(\ref{eq:parameterize}) applied to the data of table \ref{tab:frequencies2}.
In performing these fits, we set to zero the small residual phases in the $c_n$ estimates
(which are all compatible to zero, within the accuracy of the parameterizations).

For all helicities,
the resulting best fit value of the overall coefficient $b$ coincides with
the analytically known value (\ref{eq:K})
of the helicity $\pm2$ coefficient
$K = \half \left[ \sqrt \pi \, \Gamma(\tfrac 74)/\Gamma(\tfrac 54) \right]^{4/3} = 1.092\cdots$
to within a percent,
and is quite insensitive to whether one fits all 15 modes or,
for example, just modes 6 to 10.
We take this as compelling evidence that our fitting function correctly
describes the large order asymptotic behavior of spectral deviation coefficients
for helicities $\pm1$ and 0, as well as $\pm2$, 
with the same overall coefficient $K$ for all helicities.

\begin{figure}
\center
  \includegraphics[width=0.45\linewidth]{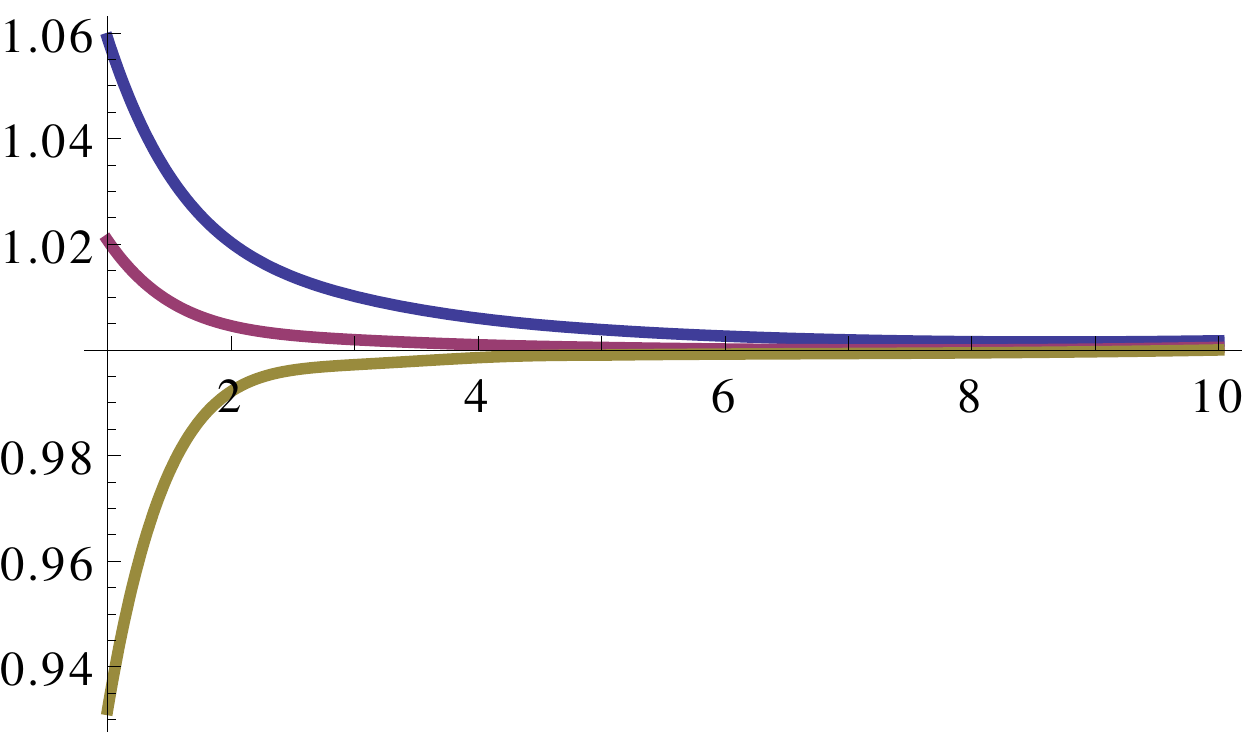}
  \put(5,62){\small $n$}
  \put(-207,130){\small $c_n^\infty/c_n$}
\vspace*{-5pt}
\caption
    {%
    Plots of the ratio $c_n^\infty/c_n$ of the large order asymptotic
    form (\ref{eq:cn-asymp}) of the spectral deviation coefficients 
    to the values (for the first ten modes) obtained from applying the parameterization 
    (\ref{eq:parameterize}) to the data of appendix \ref{app:table}.
    From top to bottom the curves correspond to helicity $\pm 2$, $\pm 1$, and $0$.
    The asymptotic form is $c_n^\infty=K(2n+|s|-1)^{4/3}$ for helicity $s$.
    }
\label{fig:cn-asympt}
\end{figure} 

If one fixes the overall coefficient $b = K$,
then the resulting best fit value for the shift $a$ is very close to an integer,
and is reasonably insensitive to the limits of the fitting range.
For helicity $\pm2$
the best fit value for the shift $a$ equals the correct answer $+1$
to within four percent.
For helicity $\pm1$,
the best fit value for the shift $a$ is quite close to zero,
somewhere in the interval $-0.002$ to $-0.03$ depending on the chosen limits
of the fitting range.
And for helicity 0,
the best fit value for the shift $a$ equals $-1$ to within a percent or two.
Given that we are only fitting data up to $n = 15$,
these results are nicely consistent with the known large order asymptotic form 
(\ref{eqn:s-alpha-asympt}) for the helicity $\pm2$ spectral deviation coefficients,
and very clearly suggest corresponding large order asymptotic forms for
helicity $\pm1$ and 0 coefficients, as reported in the introduction.
Explicitly, for helicity $s$,
$c_n \sim c_n^\infty \, [1 + \O(n^{-2})]$ with
\begin{align}
    c_n^\infty &\equiv K \, (2n + |s| - 1)^{4/3}  \,.
\label{eq:cn-asymp}
\end{align}
Figure \ref{fig:cn-asympt} shows, for each helicity,
a comparison of this asymptotic form
with our numerical estimates for spectral deviation coefficients produced by
using the functional form (\ref{eq:parameterize}) to parameterize the data of
tables \ref{tab:frequencies2}, \ref{tab:frequencies1} and \ref{tab:frequencies0}
of appendix \ref{app:table}.
The uppermost curve shows helicity $\pm 2$, the middle curve is helicity $\pm1$,
and the lower curve shows helicity 0.
Fast approach to the large order asymptotic form (\ref{eq:cn-asymp}) is manifest.
The helicity $\pm 2$ curve shown here agrees with the plot in fig.~\ref{fig:cn-asympt}, 
which used the the highly accurate $c_n$ values of table~\ref{tab:s1-values},
up to a permille.
Curiously, the approach to the asymptotic form is even faster for helicity $\pm1$ and 0 
compared to helicity $\pm2$.
For helicity $\pm1$, the deviation is only 2\% for the lowest $n=1$ mode,
and falls to half a percent or less for all higher modes.
For helicity 0, the deviation is $-7$\% for the lowest mode,
but also falls to half a percent or less for all higher modes.

\section{Planar shocks propagating in \texorpdfstring{\N{4}}{N=4} SYM plasma}\label{sec:narrow-shocks}

The general solution for the time evolution of linearized perturbations
to the metric of the AdS black brane geometry
(with fixed boundary geometry and incoming conditions at the horizon),
may be represented as a linear combination of quasinormal modes,
\begin{equation}\label{eqn:delta-g}
    \delta g(t,\x,u)
    =
    \sum_n\int \frac{d^3q}{(2\pi)^3} \>
    e^{-i\omega_n(q) t+i \q\cdot\x} \,
    A_n(\q) \, u^2 \, h_n(u;q) \, H_n \,,
\end{equation}
where $q \equiv |\q|$.
The sum runs over all quasinormal modes
(of metric perturbations)
with the symmetries of interest,
$A_n(\q)$ is the amplitude of a given mode,
and $H_n$ encodes the tensor structure of the mode,
e.g., $H_n = dx_1\otimes dx_2 + dx_2\otimes dx_1$ for helicity $\pm 2$ modes
with the indicated polarization.
Extracting a factor of $u^2$, as shown, allows one to fix the normalization
of the radial profiles $h_n(u;q)$  by requiring that they 
have a fixed boundary value, $h_n(0;q) = 1$.

The corresponding change in the expectation value of the energy-momentum
tensor of the dual QFT is then
\cite{deHaro:2000vlm}
\begin{align}\label{eqn:Tmunu-FT}
    \langle \delta T_{\mu\nu}(t,\x)\rangle
    &=
    \kappa \sum_n\int \frac{d^3q}{(2\pi)^3} \>
    e^{-i\omega_n(q) t+i \q\cdot\x} A_n(\q)\, H_{n,\mu\nu} \,,
\end{align}
where $\kappa = m^4 L^3/(4\pi G)$
with $L$ the AdS curvature scale which, elsewhere, has been set to unity.
In terms of field theory quantities,
$\kappa =\half N_c^2 \pi^2 T^4$,
where $N_c$ is the rank of the SU($N_c$) gauge group of \N{4} SYM.

Similarly, if one considers a perturbation to the equilibrium state created by
a time dependent source coupled to the QFT stress-energy tensor
(i.e., a fluctuation in the spacetime geometry of the 4D field theory),
then the induced response is given by a convolution with
the retarded stress-energy correlator,
\begin{equation}
    \langle \delta T_{\mu\nu}(t,\x)\rangle
    =
    \int \frac{d\omega \, d^3q}{(2\pi)^4} \>
    e^{-i\omega t+i\q\cdot \x} \,
    G_{\mathrm{R}}(\omega,\q)_{\mu\nu}^{\hphantom{\mu\nu}\rho\sigma} \>
    j(\omega,\q)_{\rho\sigma} \,.
\end{equation}
Quasinormal mode frequencies correspond to the poles of the
retarded Green's function \cite{Son:2002sd}, 
so evaluating the frequency integral using Cauchy's theorem yields
\begin{equation}\label{eqn:Tmunu-response}
    \langle \delta T_{\mu\nu}(t,\x)\rangle
    =
    \sum_n\int \frac{d^3q}{(2\pi)^3} \>
    e^{-i\omega_n(q) t+i\q\cdot \x} \,
    R_{n}(\q)^{\hphantom{\mu\nu}\rho\sigma}_{\mu\nu} \>
    j(\omega_n(q),\q)_{\rho\sigma} \,,
\end{equation}
where $R_n(\q)$ denotes the residue of the retarded Green's function
$G_{\mathrm{R}}(\omega,\q)$ at $\omega_n(q)$.

Both expressions (\ref{eqn:Tmunu-FT}) and (\ref{eqn:Tmunu-response})
represent the response as a sum of contributions from quasinormal modes,
and make it obvious that at sufficiently late times the response will be
dominated by those modes whose frequencies $\omega_n(\q)$ have
the smallest (in magnitude) negative imaginary parts.
Specifically,
these are long wavelength hydrodynamic modes with $q \ll T$,
together with the short wavelength modes with $q \gg T$ discussed above.
To examine qualitative features of the resulting evolution, it will be
sufficient to use the asymptotic form (\ref{eqn:disp-asymp}) of quasinormal
mode frequencies,
repeated here for convenience,
\begin{equation}\label{eqn:disp-rel-T2}
    \omega_n(q)
    =
    \pm \left[q + \half c_n \, (\pi T)^{4/3} \, q^{-1/3} \right]
    - i \tfrac {\sqrt{3}}{2} \, c_n \, (\pi T)^{4/3} \, q^{-1/3}
    +\O\big(T^3 q^{-2}\big) \,,
\end{equation}
which for low order modes, as discussed in sec.~\ref{sec:numerics},
is already quite accurate at intermediate values of $q/T$.
(Explicit values of the coefficients $c_n$ are given in tables
\ref{tab:s1-values}, \ref{tab:s1-values-spin1} and \ref{tab:s1-values-spin0}.)

\subsection{Fine structures outlive coarse}

Consider a metric perturbation $\delta g$,
represented in the form (\ref{eqn:delta-g}),
which at time $t = 0$ has rapid spatial variation and
a spatial Fourier transform concentrated near
some wavevector $\q_0$ with $|\q_0| \gg T$.
The asymptotic form (\ref{eqn:disp-rel-T2}) implies that
the characteristic relaxation time of such an excitation
will be of order $\tau(q_0) \equiv q_0^{1/3} (\pi T)^{-4/3}$,
with higher modes (larger $n$) damping faster than lower modes.
At times comparable or larger than this relaxation time,
dominant contributions will come from the $n{=}1$ mode with
wavenumbers near $\q_0$.%
\footnote
    {%
    More precisely, dominant contributions can come from the lowest mode
    in each helicity channel.
    For simplicity, we ignore the presence of multiple helicity channels
    in the following qualitative discussion.
    }
Provided the perturbation is sufficiently small,
so a linearized treatment is valid, there is no mode-mixing populating
other ranges of wavevector.
The resulting late-time energy-momentum tensor (\ref{eqn:Tmunu-FT})
is then well described by just the $n\,{=}\,1$ contribution,
\begin{equation}\label{eqn:Tmunu}
    \langle \delta T_{\mu\nu}\rangle
    =
    \kappa \int \frac{d^3q}{(2\pi)^3} \>
    e^{i\phi(\q)} \, \widetilde A(\q) \, H_{1,\mu\nu} \,,
\end{equation}
with a damped amplitude
\begin{equation}
    \widetilde A(\q)
    \equiv
    A_1(\q) \,
    \exp\big[{-t \,\tfrac{\sqrt 3}{2} \, c_1 \, (\pi T)^{4/3} \, q^{-1/3}} \big]\,,
\label{eq:damping}
\end{equation}
and rapidly varying phase
\begin{equation}
    \phi(\q)
    \equiv
    \q\cdot\x
    - \big[ q + \half \, c_1 \, (\pi T)^{4/3} \, q^{-1/3} \big] \, t \,.
\end{equation}
If one asks when a disturbance, initially localized near
$\x = 0$ at time 0, will reach a distant point $\x$, the dominant
contributions to the integral (\ref{eqn:Tmunu}) come from the neighborhood
of the stationary phase point where $\nabla \phi(\q) = 0$.
(provided $\widetilde A(\q)$ is slowly varying on the scale of $|\x|^{-1}$).
This yields the standard result that disturbances localized
in wavenumber near $\q = \q_0$ travel at the group velocity,
\begin{equation}
    {\bf v}_{\rm g}(\q_0)
    \equiv \nabla \Re(\omega(\q_0))
    = v_{\rm g}(q_0) \, \hat \q_0 \,,
    \qquad
    v_{\rm g}(q_0) \sim
     1 - \tfrac 16 \, c_1 \, (\pi T/q_0)^{4/3} \,,
\label{eq:vg}
\end{equation}
and hence arrive at position $\x = d \, \hat \q_0$ at time
$
    t = d / v_{\rm g}(q_0)
$.

The damping (\ref{eq:damping}) decreases monotonically with
increasing wavenumber, while the group velocity (\ref{eq:vg})
increases monotonically,
asymptotically approaching the speed of light.
Hence, shorter wavelength features attenuate slower,
and propagate faster, than longer wavelength features.
For disturbances with a significant range of wavenumbers,
the overall evolution will reflect a combination of the
wavenumber dependent damping (\ref{eq:damping}) and the
dispersive propagation (\ref{eq:vg}).

\subsection{Planar shocks at late times}

The evolution of planar shocks in a strongly coupled $\N{4}$ SYM plasma
provides an interesting, concrete illustration of the above features.
At zero temperature, planar shocks (composed of helicity 0 perturbations)
move at the speed of light with no dispersion or damping.
For a shock propagating in, say, the $+\hat x_3$ direction
with an arbitrary longitudinal energy density profile $\kappa \, h(x_3)$,
stress-energy components at time $t$ are
\begin{equation}
    \delta T^{00}(t,\x) = \delta T^{03}(t,\x) = \delta T^{33}(t,\x)
    =
    \kappa \, h(x_3{-}t) \,,
\end{equation}
with all other components vanishing.
Equivalently,
\begin{equation}
    \langle \delta T_{\mu\nu}\rangle
    =
    \kappa \int \frac{dq_3}{2\pi} \> e^{iq_3(x_3-t)} \, A(q_3)\>
    (dx_3{-}dt)_\mu \, (dx_3{-}dt)_\nu \,,
\end{equation}
with $A(q_3)$ the Fourier transform of $h(x_3)$.
The dual geometry is an exact analytic solution of Einstein's equations
\cite{Chesler:2010bi}.
Analogous planar stress-energy perturbations with helicity
$\pm1$ or $\pm2$ tensor structures correspond to solutions of Einstein's
equations linearized about $AdS_5$, but analytic solutions at the
non-linear level are not known.

We are interested in the modification in the evolution of planar shocks
induced by the presence of a background thermal plasma.%
\footnote
    {%
    Ensuring that initial data for Einstein's equations
    are consistent with initial value constraints can be tricky.
    But in infalling coordinates the identification of unconstrained
    initial data is easy, and it is consistent to simply add a
    background energy density \cite{Chesler:2010bi,Chesler:2013lia}
    and start the evolution at time $t = 0$.
    }
We assume that the amplitude of the shock is sufficiently small so that
a linearized treatment is adequate.
And, for simplicity, we assume that the perturbation has a single
tensor structure corresponding to either helicity 0, $\pm1$, or $\pm 2$.

Planarity of the shock implies that the expression (\ref{eqn:Tmunu-FT})
for the stress-energy (at times $t \ge 0$) simplifies to a one-dimensional
Fourier transform,
\begin{equation}
    \langle \delta T_{\mu\nu}(t,x_3)\rangle
    =
    \kappa \sum_n\int \frac{dq_3}{2\pi} \>
    e^{-i\omega_n(q_3) t+i q_3 x_3} \,
    A_n(q_3) \, H_{\mu\nu} \,,
\label{eq:multimodeT}
\end{equation}
with the coefficients $\{ A_n(q_3) \}$ characterizing
the chosen shock profile at time $t=0$.

As discussed in the previous subsection,
since large-$q$ modes experience minimal damping,
rapidly varying features in the longitudinal profile of the shock
will outlive more slowly varying features.
A particularly clear picture emerges for narrow shocks.
A shock with narrow width $w \ll 1/T$ will have a broad Fourier spectrum 
extending from small wavenumbers at least up to $|q_3| \sim 1/w$.
More precisely, the breadth of the Fourier spectrum reflects
the (inverse) scale of variation of the sharpest spatial features.
As a coherent superposition of many different wavevectors,
small differences in the propagation of different wavenumbers will
imprint themselves  on the evolution of the shock profile.
In particular, since the speed of propagation approaches the speed of light
as $|q_3|\to\infty$, 
contributions from the highest-wavenumber modes will accumulate very
near the light cone, forming an increasingly sharp leading edge.
These sharp features will persist longer than contributions from
lower wavenumbers (except for very small $q$ hydrodynamic modes)
which attenuate more quickly.

To illustrate this explicitly with simple examples,
we consider perturbations which are dominated by the lowest
quasinormal mode (of a given helicity), so that
\begin{equation}
    \langle \delta T_{\mu\nu}(t,x_3)\rangle
    =
    \kappa \int \frac{dq_3}{2\pi} \> e^{-i\omega_1(q_3) t+i q_3 x_3} \,
    A_1(q_3) \, H_{\mu\nu} \,.
\label{eq:onemodeT}
\end{equation}
One may either regard this as fine-tuning the initial data,
or the result of starting with a more general perturbation and
waiting to sufficiently late times where higher modes are
small compared to the lowest mode.
For simplicity, we include only modes with $\Re(q_3 \, \omega_1(q_3)) > 0$
in the perturbation (\ref{eq:onemodeT}), with no corresponding
contributions from the reflected modes with frequency $-\omega_1(q_3)^*$;
this means that we are focusing on right-moving perturbations.
We compare three different longitudinal profiles,%
\footnote
    {%
    Each of these profiles should be regarded as multiplied by
    some small parameter $\epsilon$.
    But, as the entire analysis is linearized, we shall omit
    writing $\epsilon$ explicitly.
    }
\begin{equation}
    A_1^{\rm Gauss}(q)
    =
    e^{-\frac 12 \sigma^2 q^2} \,,
    \qquad
    A_1^{\rm Blob}(q)
    =
    \frac{2 J_1(\sigma q)}{\sigma q} \,,
    \qquad
    A_1^{\rm Step}(q)
    =
    \frac{\sin(\sigma q)}{\sigma q} \,.
\end{equation}
These are Fourier transforms of position space profiles
$h(x_3) = \int dq/(2\pi) \> e^{i q x_3} \, A_1(q)$
which are, respectively, a Gaussian,
a semicircular ``blob,''
and a ``top-hat'' step function,
each normalized to unit area:
\begin{equation}\label{eqn:position-space-profiles}
    h^{\rm Gauss}(x_3)
    =
    \frac{e^{-\frac 12 {x_3^2}/\sigma^2}}{\sqrt{2\pi} \, \sigma} \,,
    \quad
    h^{\rm Blob}(x_3)
    =
    \frac{2\sqrt{\sigma^2{-}x_3^2}}{\pi\sigma^2} \, \Theta(\sigma^2{-}x_3^2)
    \,,
    \quad
    h^{\rm Step}(x_3)
    =
    \frac{\Theta(\sigma^2{-}x_3^2)}{2\sigma} \,.
\end{equation}
We choose $\sigma=1/10$ for $h^{\rm Gauss}$,
and $\sigma=1/5$ for $h^{\rm Blob}$ and $h^{\rm Step}$.

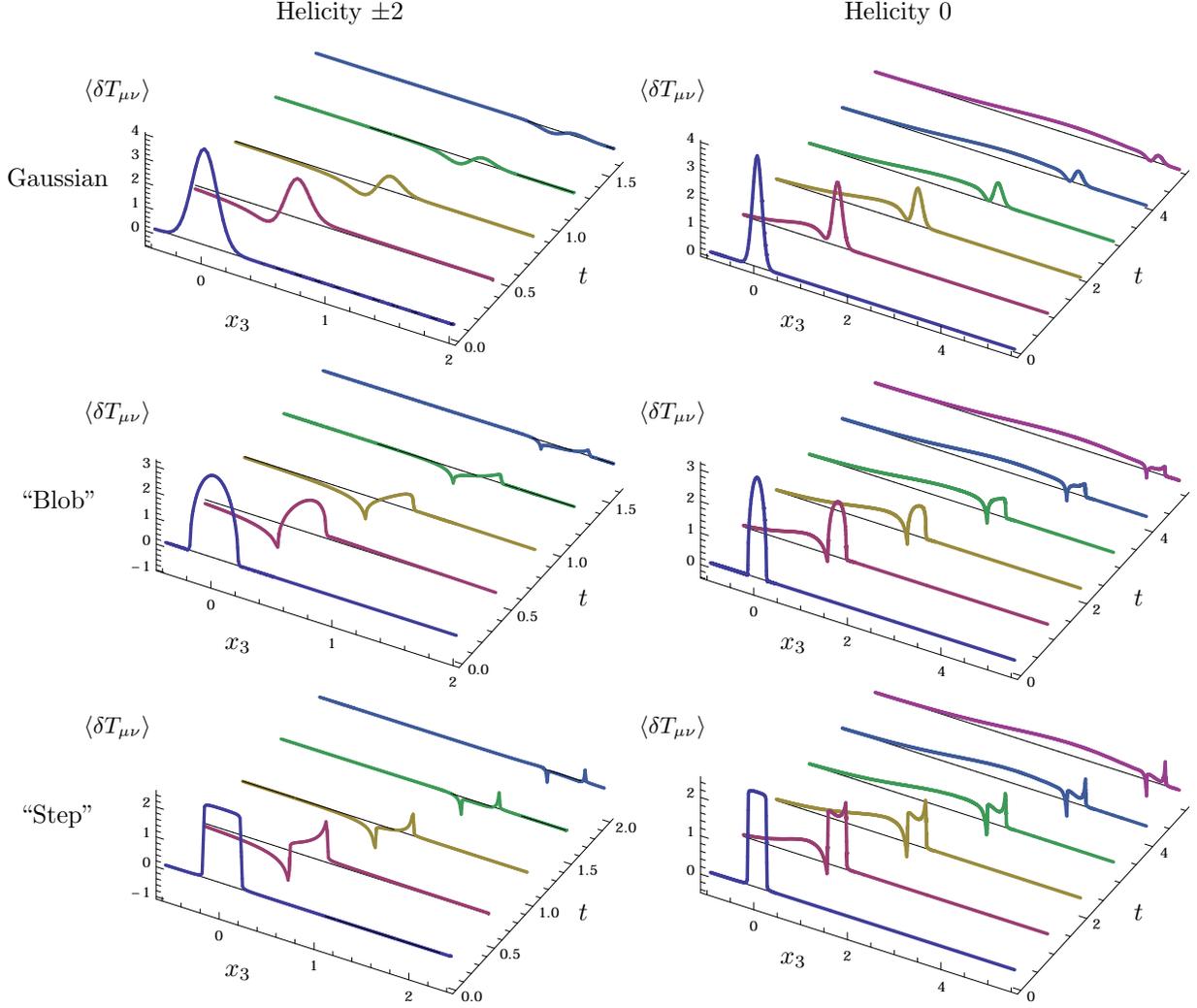
\begin{figure}
  \hfill
  \includegraphics[width=0.42\linewidth]{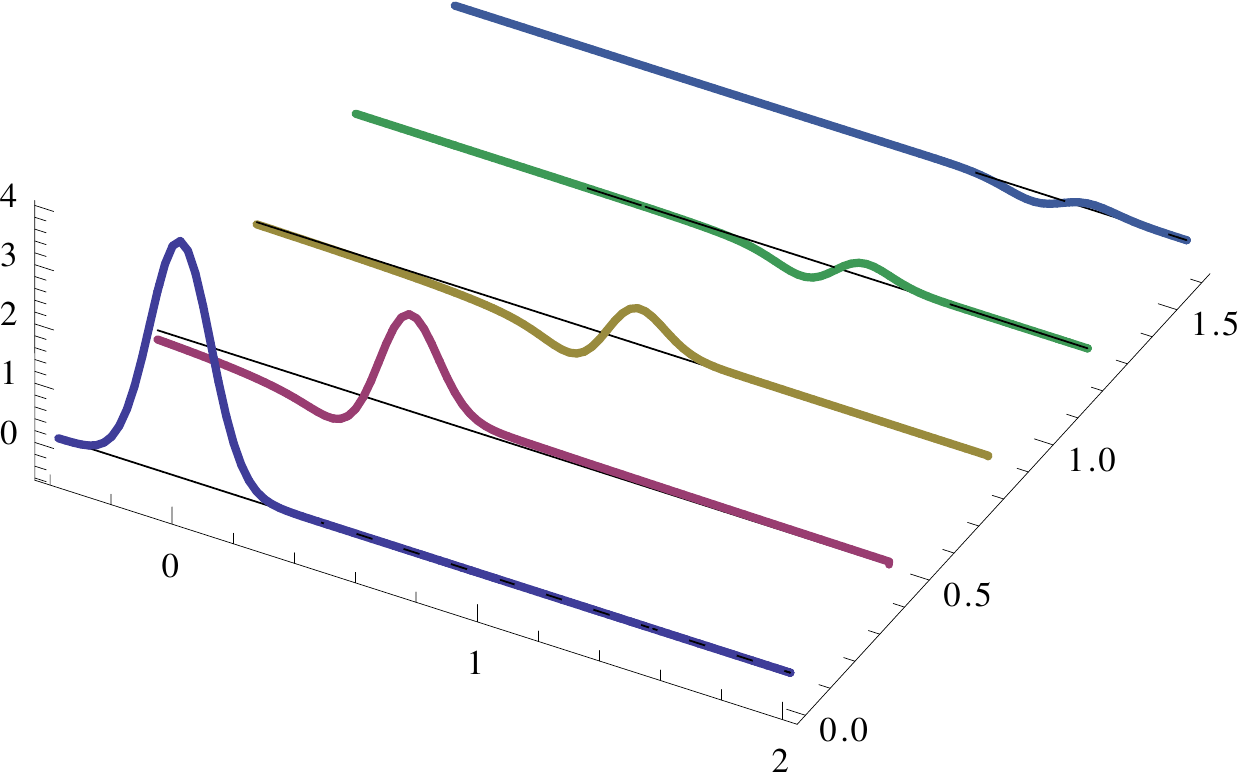} 
  \put(-140,135){Helicity $\pm2$}
  \put(-245,70){Gaussian}
  \put(-160,15){\small $x_3$}
  \put(-22,32){\small $t$}
  \put(-215,105){\footnotesize $\langle \delta T_{\mu\nu}\rangle$}
\qquad
  \includegraphics[width=0.42\linewidth]{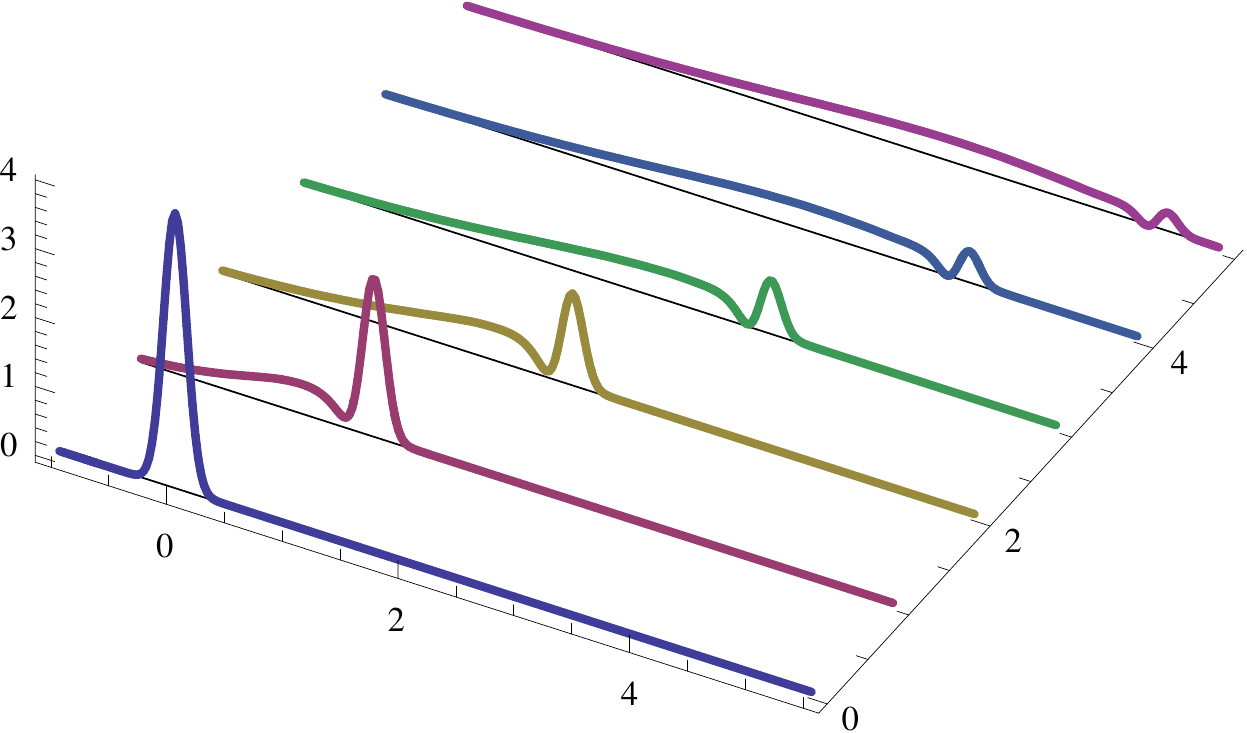}
  \put(-135,135){Helicity $0$}
  \put(-160,15){\small $x_3$}
  \put(-22,32){\small $t$}
  \put(-215,105){\footnotesize $\langle \delta T_{\mu\nu}\rangle$}
\\
  \hfill
  \includegraphics[width=0.42\linewidth]{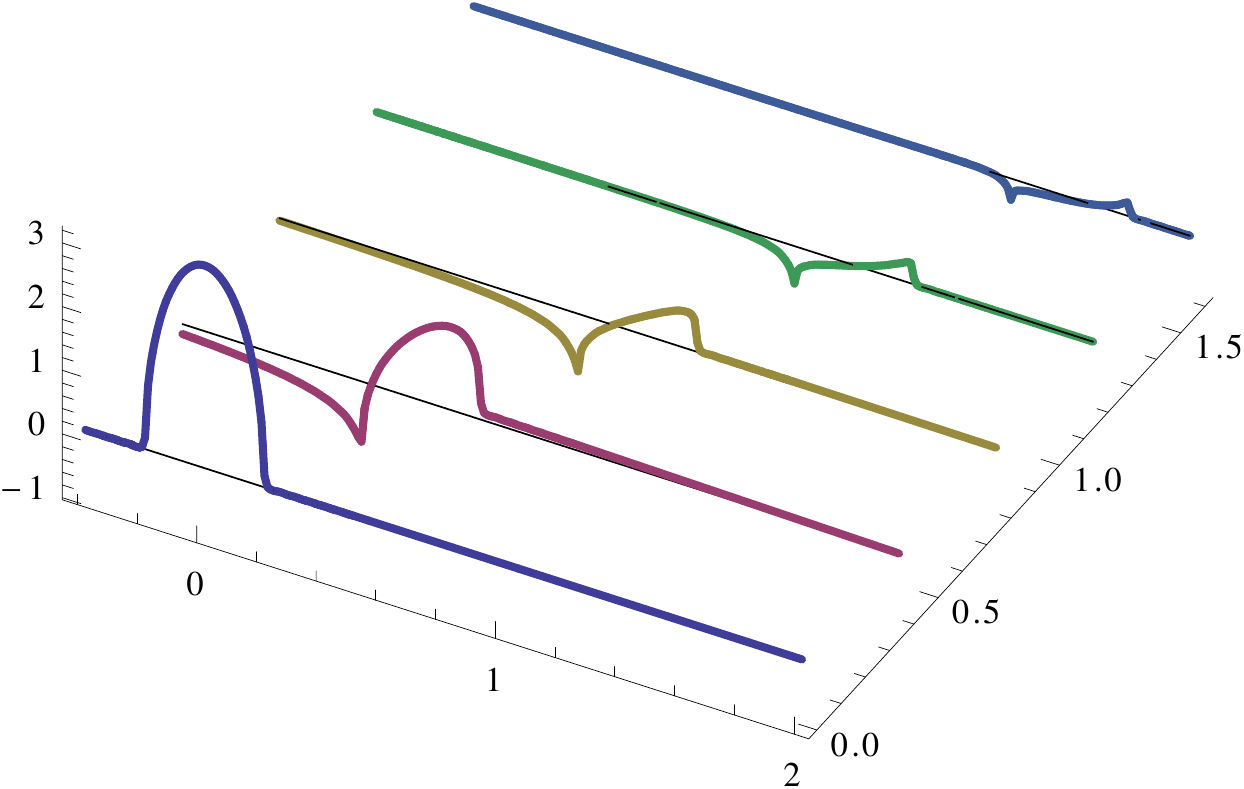} 
  \put(-240,70){``Blob''}
  \put(-160,15){\small $x_3$}
  \put(-22,32){\small $t$}
  \put(-215,105){\footnotesize $\langle \delta T_{\mu\nu}\rangle$}
\qquad
  \includegraphics[width=0.42\linewidth]{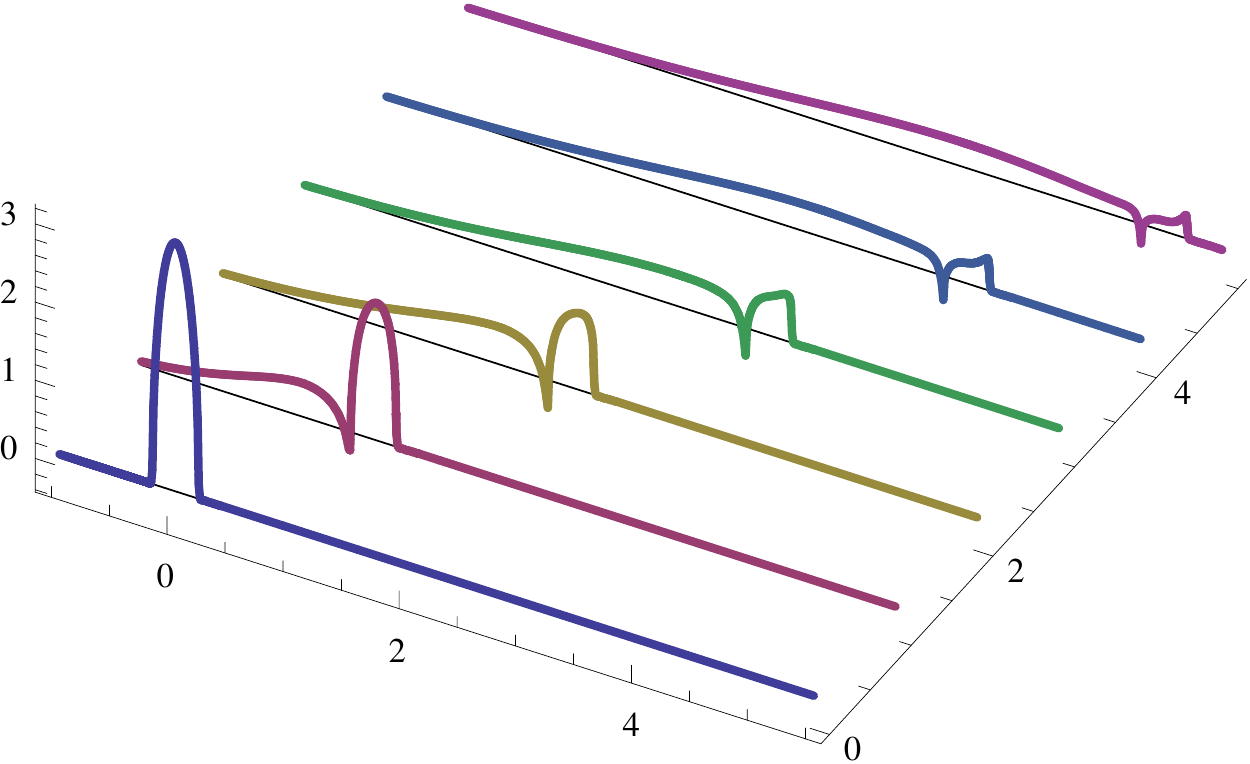} 
  \put(-160,15){\small $x_3$}
  \put(-22,32){\small $t$}
  \put(-215,105){\footnotesize $\langle \delta T_{\mu\nu}\rangle$}
\\
  \hfill
  \includegraphics[width=0.42\linewidth]{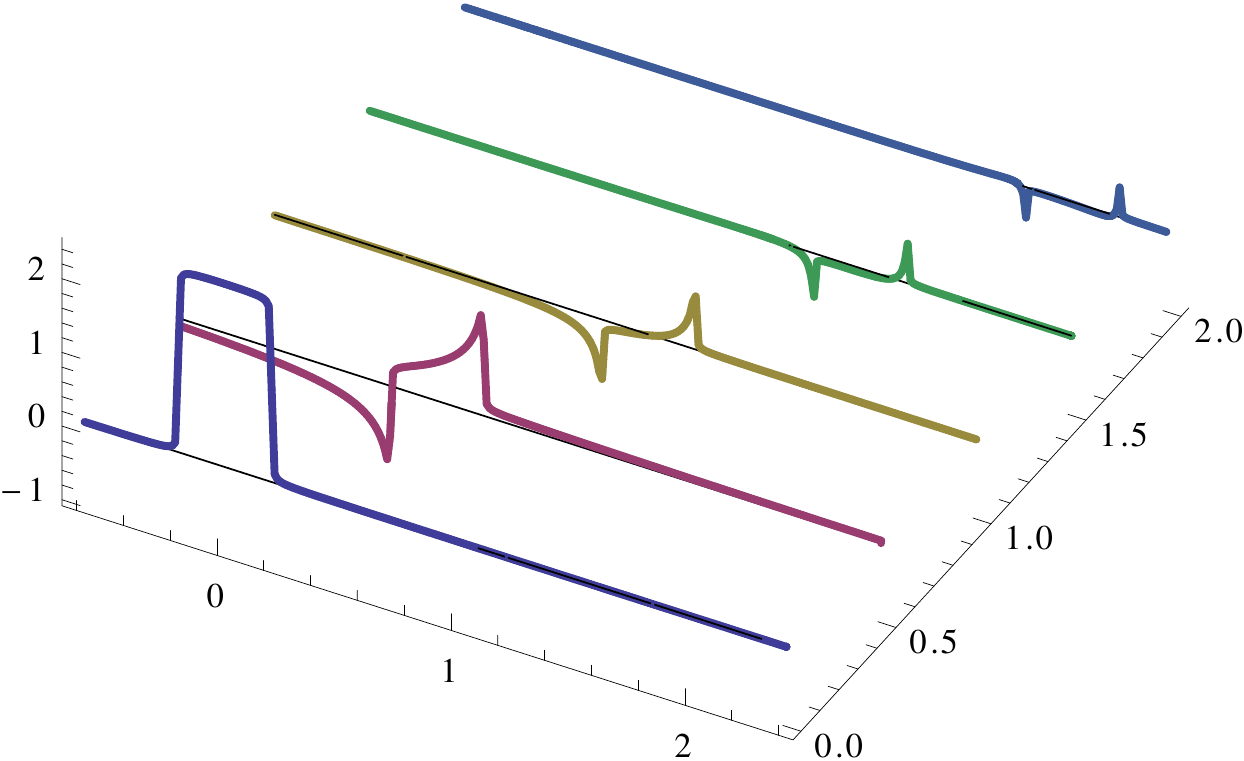} 
  \put(-240,70){``Step''}
  \put(-160,10){\small $x_3$}
  \put(-22,32){\small $t$}
  \put(-215,105){\footnotesize $\langle \delta T_{\mu\nu}\rangle$}
\qquad
  \includegraphics[width=0.42\linewidth]{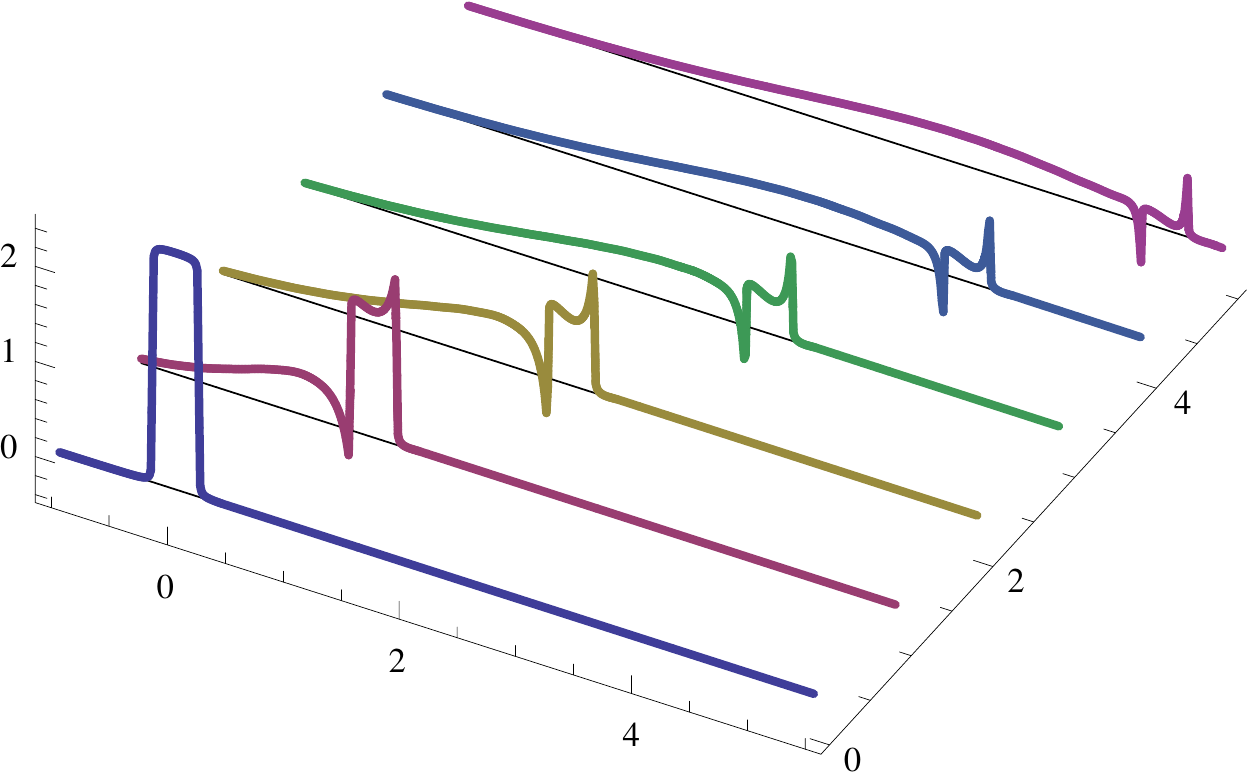}
  \put(-160,15){\small $x_3$}
  \put(-22,32){\small $t$}
  \put(-215,105){\footnotesize $\langle \delta T_{\mu\nu}\rangle$}
\caption
    {%
    Time evolution of helicity $\pm2$ perturbations [left] and
    helicity 0 perturbations [right] for each of the three initial profiles
    (\ref{eqn:position-space-profiles}):
    Gaussian, ``Blob,'' and ``Step'' [top to bottom].
    Snapshots are taken at times
    $t= 0,\frac{1}{2},1,\frac 32,2$ for helicity $\pm 2$,
    and at 
    $t= 0,1,2,3,4,5$ for helicity 0;
    the different time scales reflect the faster damping of helicity $\pm 2$
    perturbations.
    For the helicity-$2$ perturbations one sees that the longest surviving
    features are associated with the steepest portions of the initial
    profile.
    This is especially apparent with the compactly-supported
    ``blob'' and ``step'' profiles.
    For helicity $0$ there is, in addition, a visible slowly
    varying contribution from the hydrodynamic sound mode.
    }
\label{fig:shocks}
\end{figure}

For the dispersion relation $\omega_1(q)$ we construct a cubic spline
interpolating function from
the numerical results of sec.~\ref{sec:numerics} for
low and intermediate wavenumbers,
which smoothly connects to the large-$q$
asymptotics of eq.~(\ref{eqn:disp-rel-T2}).
Since the large-$q$ asymptotic form is already quite accurate for
intermediate values of $q$, this procedure is straightforward.%
\footnote
    {%
    In fact, just na\"ively using the large-$q$ asymptotic form for all $q$,
    with a simple IR cut-off to regularize the $q^{-1/3}$ term,
    only mildly changes the helicity $\pm2$ results.
    For helicity 0, such a naive approach omits contributions from
    the $q \to 0$ hydrodynamic mode.
    }
Figure~\ref{fig:shocks} shows
plots of the resulting time evolution for perturbations
with each of the above profiles,
for the helicity $\pm2$ tensor structure
$dx_1 \otimes dx_2$ as well as the helicity 0 structure
$(dx_3 {-} dt) \otimes (dx_3 {-} dt)$.
Comparing the plots, one clearly sees the stronger damping of
helicity $\pm2$ fluctuations relative to helicity $0$,
due to the larger values of the dispersive coefficient
$c_1$ (cf.\ tables \ref{tab:s1-values} and \ref{tab:s1-values-spin0}).
For the helicity $\pm2$ perturbations, shown on the
left side of the figure,
the longest surviving features
are associated with the steepest portions of the initial profile.
This is especially apparent for the ``blob'' and ``step'' profiles
which have compact support.
At late times, one sees upward ``spikes'' which evolve from the portion
of the initial profile with large negative gradient, and downward spikes
evolving from the steeply rising part of the initial profile.

The helicity 0 evolution, shown on the right-hand side of the figure,
shows similar sharp high-$q$ features but with the addition of
a slowly varying hydrodynamic (sound) contribution which moves
slower than the leading features and gradually broadens.
(The speed of sound in the conformal $\N{4}$ plasma is $1/\sqrt 3$
\cite{Kovtun:2005ev}.)
Hence, as time increases there is an increasingly clear separation
between the attenuating high-$q$ and low-$q$ remnants.

The helicity 0 planar shocks have a conserved energy (and linear momentum).
At late times, the energy and momentum of the shock is entirely carried
by the $q\to0$ hydrodynamic contribution;
for the profiles of fig.~\ref{fig:shocks},
the upward and downward high-$q$ spikes cancel each other upon integration.
More generally, the long-lived fine structure carries little net energy or momentum.
This might seem odd, since high momentum quasiparticles (or short wavelength
waves) in other contexts can transport energy and momentum.
But stress-energy quasinormal modes in strongly coupled (and large $\Nc$)
SYM plasma are perturbations in which the energy density, momentum density,
and/or stress of the fluid vary sinusoidally (for non-zero $\q$) and hence
unavoidably vanish upon integration.
This is fundamentally different from, say, an electromagnetic wave
in vacuum in which the EM field varies sinusoidally but the energy density
is quadratic in the wave amplitude and always positive.

\section{Discussion}\label{sec:discussion}

We have extended and clarified previous work on
quasinormal mode frequencies for metric perturbations of
AdS-Schwarzschild black branes, or equivalently stress-energy
perturbations in strongly coupled $\N{4}$ SYM plasma.
The large wavenumber asymptotic behavior
has the universal form (\ref{eqn:disp-rel-T2}),
in all helicity channels,
with mode-dependent spectral deviation coefficients shown in
tables~\ref{tab:s1-values}, \ref{tab:s1-values-spin1} and \ref{tab:s1-values-spin0}.
The relaxation rate of short wavelength quasinormal modes
vanishes asymptotically as $T^{4/3} q^{-1/3}$.
We find that the large-$q$ asymptotic form (\ref{eqn:disp-rel-T2})
of quasinormal mode frequencies
agrees well with the exact values (for low order modes)
already at rather moderate values of $q/T$,
and thus provides a good approximation over a wide range of scales.
The spectral deviation coefficients of high order modes approach the
asymptotic form (\ref{eq:cn-asymp}) but
the coefficients of low order modes deviate from this simple expression.

In the strongly coupled SYM plasma it is noteworthy that
hydrodynamic fluctuations are not the only arbitrarily long-lived
excitations.
The asymptotic vanishing of relaxation rates implies that
there are two types of long-lived propagating excitations:
long-wavelength sound waves, moving at $v_{\rm s} = c/\sqrt 3$,
and short-wavelength fluctuations with group velocities
asymptotically approaching $c$.
As vividly seen in figure~\ref{fig:shocks},
illustrating examples of planar shock propagation,
the decreasing attenuation plus increasing group velocity 
(as the wavevector $q$ increases)
combine to produce sharp, long-lived ``spikes'' which evolve
from the most rapidly varying parts of an initial waveform.
This same phenomena is evident in the study \cite{Chesler:2011nc}
by Chesler, Ho and Rajagopal
of the ``cyclotron radiation'' produced by circular stirring of
an SYM plasma (see fig.~3 of ref.~\cite{Chesler:2011nc}).
Whether such long-lived ``fine structure'' could have observable
phenomenological consequences in heavy ion collisions is an interesting
open question.

There has been considerable discussion in the literature
\cite{%
    Steineder:2012si,
    Balasubramanian:2010ce,
    Balasubramanian:2011ur,
    Galante:2012pv,
    Steineder:2013ana,
    Stricker:2013lma,
    Stricker:2014cma}
regarding ``top-down'' thermalization in strongly coupled SYM, as
compared with ``bottom-up'' thermalization at weak coupling \cite{Baier:2000sb}.
Evidence suggesting a top-down picture of thermalization
at strong coupling comes from the decreasing relaxation times
of quasinormal modes as the mode number increases at fixed wavenumber,
and related observables probing similar physics.%
\footnote
    {%
    In particular, calculations
    of the first finite coupling corrections to quasinormal mode frequencies
    \cite{Stricker:2013lma,Stricker:2014cma}
    suggested that,
    for intermediate values of the 't Hooft coupling,
    damping rates (for fixed wavenumber) cease to monotonically increase
    with increasing mode number,
    implying a change in character of relaxation processes as the
    coupling decreases from asymptotically large values.
    More recent work \cite{Waeber:2015oka,Grozdanov:2016vgg}
    finds that this behavior was a consequence of extrapolating 
    next-to-leading order results outside their regime of validity,
    with more complete calculations showing no sign of any change
    in monotonicity with mode order as the coupling decreases.
    }
However, interpreting this as top-down thermalization is,
in our view, conflating the dephasing or decoherence of highly
virtual off-shell excitations ($|\omega|^2 \gg q^2)$,
with relaxation of high momentum but near on-shell excitations
($|\omega-q|^2 \ll q^2$).
It is these latter excitations, corresponding to large wavenumber but
low order quasinormal modes, which should be considered when discussing
top-down versus bottom-up thermalization.
And these hard but low virtuality excitations thermalize slowly
at both weak and strong coupling.%
\footnote
    {%
    This same point concerning
    the qualitative difference in thermalization
    properties of highly virtual versus on-shell but large $q$
    fluctuations was observed and discussed much earlier in the
    seminal paper \cite{CaronHuot:2011dr} by Caron-Huot, Chesler and Teaney.
    }

Finally, we limited our attention to metric
(or stress-energy) perturbations.
We expect quasinormal mode frequencies for perturbations
in other supergravity fields to have the same universal
asymptotic form (\ref{eqn:disp-rel-T2}),
but it would be worthwhile to demonstrate (or disprove) this explicitly,
especially for fermionic fluctuations.
We leave such questions for future work.

\begin{acknowledgments}

We thank Alex Buchel, Paul Chesler, and Andrei Starinets
for helpful conversations.
This work was supported, in part, by the U.S.~Department of Energy
under Grant No.~DE-SC0011637.

\end{acknowledgments}

\clearpage

\appendix

\section{Tabulated results}\label{app:table}

\begin{table}[ht]
\setlength{\tabcolsep}{5pt}
\def\i{\,i}
\footnotesize
\begin{tabular}{c||c|c|c|c|c}
\hline
\hline
\multicolumn{6}{c}{helicity $\pm2$}\vphantom{\Large$\strut$}\\
\hline
\hline
$n$ & $q=10$ & $q=20$ & $q=40$ & $q=80$ & $q=160$\\
\hline
 1 & $1.05863-1.75039\i$ & $0.82995-1.41031\i$ & $0.65513-1.12603\i$ & $0.51880-0.89583\i$ & $0.41139-0.71168\i$ \\
 2 & $2.20415-3.49828\i$ & $1.71646-2.86252\i$ & $1.34867-2.29987\i$ & $1.06566-1.83424\i$ & $0.84425-1.45862\i$ \\
 3 & $3.52349-5.37997\i$ & $2.73704-4.47509\i$ & $2.14189-3.62096\i$ & $1.68857-2.89602\i$ & $1.33636-2.30553\i$ \\
 4 & $4.97581-7.33629\i$ & $3.87015-6.20226\i$ & $3.01901-5.05684\i$ & $2.37469-4.05696\i$ & $1.87734-3.23373\i$ \\
 5 & $6.52930-9.33352\i$ & $5.09962-8.01283\i$ & $3.96941-6.58524\i$ & $3.11555-5.30074\i$ & $2.46028-4.23071\i$ \\
 6 & $8.15996-11.3527\i$ & $6.41236-9.88471\i$ & $4.98539-8.18992\i$ & $3.90521-6.61561\i$ & $3.08037-5.28757\i$ \\
 7 & $9.85002-13.3833\i$ & $7.79737-11.8019\i$ & $6.06096-9.85842\i$ & $4.73923-7.99264\i$ & $3.73400-6.39761\i$ \\
 8 & $11.5864-15.4192\i$ & $9.24519-13.7526\i$ & $7.19119-11.5808\i$ & $5.61417-9.42475\i$ & $4.41841-7.55557\i$ \\
 9 & $13.3592-17.4572\i$ & $10.7476-15.7280\i$ & $8.37186-13.3490\i$ & $6.52726-10.9061\i$ & $5.13137-8.75720\i$ \\
 10& $15.1614-19.4954\i$ & $12.2977-17.7217\i$ & $9.59927-15.1564\i$ & $7.47616-12.4320\i$ & $5.87107-9.99894\i$ \\
 11& $16.9874-21.5328\i$ & $13.8892-19.7288\i$ & $10.8701-16.9972\i$ & $8.45891-13.9981\i$ & $6.63596-11.2778\i$ \\
 12& $18.8329-23.5687\i$ & $15.5171-21.7457\i$ & $12.1812-18.8667\i$ & $9.47378-15.6008\i$ & $7.42475-12.5912\i$ \\
 13& $20.6946-25.6031\i$ & $17.1769-23.7697\i$ & $13.5299-20.7609\i$ & $10.5193-17.2372\i$ & $8.23632-13.9368\i$ \\
 14& $22.5701-27.6357\i$ & $18.8647-25.7988\i$ & $14.9136-22.6763\i$ & $11.5940-18.9042\i$ & $9.06968-15.3127\i$ \\
 15& $24.4571-29.6665\i$ & $20.5772-27.8314\i$ & $16.3299-24.6099\i$ & $12.6966-20.5994\i$ & $9.92396-16.7170\i$ \\
\hline\hline
\end{tabular}
\caption
    {%
    Dispersive corrections $\omega_n(q)-q$ of the first fifteen helicity $\pm 2$ quasinormal mode frequencies,
    for wavenumbers $q = 10$, 20, 40, 80 and 160.
    Both frequencies and wavenumbers are in units of $\pi T$.
    Values were obtained by numerically solving
    eqn.~(\ref{eqn:radial-eq});
    all digits shown are accurate.
    }
\label{tab:frequencies2}
\end{table}

\begin{table}[ht!]
\setlength{\tabcolsep}{5pt}
\def\i{\,i}
\footnotesize
\begin{tabular}{c||c|c|c|c|c}
\hline
\hline
\multicolumn{6}{c}{helicity $\pm1$}\vphantom{\Large$\strut$}\\
\hline
\hline
$n$ & $q=10$ & $q=20$ & $q=40$ & $q=80$ & $q=160$\\
\hline
 1& $0.58720-1.15903\i$ & $0.48382-0.88383\i$ & $0.39014-0.69032\i$ & $0.31165-0.54439\i$ & $0.24799-0.43098\i$ \\
 2& $1.58123-2.81103\i$ & $1.26570-2.21333\i$ & $1.00771-1.75182\i$ & $0.80074-1.38891\i$ & $0.63583-1.10191\i$ \\
 3& $2.79183-4.65861\i$ & $2.20774-3.75118\i$ & $1.74488-2.99701\i$ & $1.38179-2.38510\i$ & $1.09564-1.89507\i$ \\
 4& $4.16276-6.60981\i$ & $3.27674-5.42911\i$ & $2.57610-4.37611\i$ & $2.03409-3.49506\i$ & $1.61075-2.78092\i$ \\
 5& $5.65418-8.61660\i$ & $4.45197-7.20641\i$ & $3.48687-5.85979\i$ & $2.74601-4.69693\i$ & $2.17175-3.74256\i$ \\
 6& $7.23681-10.6529\i$ & $5.71811-9.05590\i$ & $4.46762-7.42824\i$ & $3.51003-5.97618\i$ & $2.77255-4.76890\i$ \\
 7& $8.88910-12.7042\i$ & $7.06275-10.9583\i$ & $5.51134-9.06690\i$ & $4.32085-7.32229\i$ & $3.40884-5.85199\i$ \\
 8& $10.5953-14.7627\i$ & $8.47545-12.8998\i$ & $6.61244-10.7645\i$ & $5.17446-8.72715\i$ & $4.07739-6.98577\i$ \\
 9& $12.3438-16.8240\i$ & $9.94725-14.8700\i$ & $7.76630-12.5119\i$ & $6.06770-10.1843\i$ & $4.77567-8.16543\i$ \\
10& $14.1260-18.8855\i$ & $11.4705-16.8615\i$ & $8.96890-14.3018\i$ & $6.99800-11.6883\i$ & $5.50164-9.38705\i$ \\
11& $15.9354-20.9461\i$ & $13.0384-18.8684\i$ & $10.2167-16.1280\i$ & $7.96319-13.2348\i$ & $6.25361-10.6473\i$ \\
12& $17.7671-23.0051\i$ & $14.6455-20.8867\i$ & $11.5065-17.9852\i$ & $8.96143-14.8197\i$ & $7.03017-11.9434\i$ \\
13& $19.6174-25.0620\i$ & $16.2868-22.9133\i$ & $12.8353-19.8690\i$ & $9.99107-16.4398\i$ & $7.83008-13.2730\i$ \\
14& $21.4833-27.1168\i$ & $17.9582-24.9457\i$ & $14.2005-21.7757\i$ & $11.0507-18.0920\i$ & $8.65231-14.6338\i$ \\
15& $23.3624-29.1695\i$ & $19.6561-26.9822\i$ & $15.5995-23.7021\i$ & $12.1389-19.7737\i$ & $9.49590-16.0240\i$ \\
\hline\hline
\end{tabular}
\caption
    {%
    Dispersive corrections $\omega_n(q)-q$ of the first fifteen helicity $\pm 1$ quasinormal mode frequencies,
    for wavenumbers $q = 10$, 20, 40, 80 and 160.
    Both frequencies and wavenumbers are in units of $\pi T$.
    Values were obtained by numerically solving
    eqn.~(\ref{eqn:radial-spin1});
    all digits shown are accurate.
    }
\label{tab:frequencies1}
\end{table}

\begin{table}[ht]
\setlength{\tabcolsep}{5pt}
\def\i{\,i}
\footnotesize
\begin{tabular}{c||c|c|c|c|c}
\hline
\hline
\multicolumn{6}{c}{helicity $0$}\vphantom{\Large$\strut$}\\
\hline\hline
$n$ & $q=10$ & $q=20$ & $q=40$ & $q=80$ & $q=160$\\
\hline
 1& $0.22331-0.53674\i$ & $0.20262-0.39493\i$ & $0.16745-0.30347\i$ & $0.13483-0.23771\i$ & $0.10760-0.18768\i$ \\
 2& $1.03696-2.01064\i$ & $0.85842-1.55109\i$ & $0.69082-1.21637\i$ & $0.55112-0.96076\i$ & $0.43829-0.76108\i$ \\
 3& $2.13166-3.77754\i$ & $1.71860-2.99339\i$ & $1.36816-2.37364\i$ & $1.08667-1.88324\i$ & $0.86267-1.49450\i$ \\
 4& $3.41657-5.68700\i$ & $2.72292-4.60793\i$ & $2.15230-3.68885\i$ & $1.70369-2.93794\i$ & $1.35053-2.33504\i$ \\
 5& $4.84199-7.67141\i$ & $3.84399-6.34082\i$ & $3.02311-5.12263\i$ & $2.38595-4.09505\i$ & $1.88883-3.25950\i$ \\
 6& $6.37289-9.69513\i$ & $5.06377-8.15857\i$ & $3.96864-6.65069\i$ & $3.12398-5.33655\i$ & $2.46988-4.25403\i$ \\
 7& $7.98371-11.7388\i$ & $6.36836-10.0382\i$ & $4.98063-8.25598\i$ & $3.91141-6.65002\i$ & $3.08852-5.30917\i$ \\
 8& $9.65577-13.7918\i$ & $7.74630-11.9633\i$ & $6.05280-9.92562\i$ & $4.74359-8.02619\i$ & $3.74100-6.41796\i$ \\
 9& $11.3754-15.8482\i$ & $9.18783-13.9218\i$ & $7.18006-11.6495\i$ & $5.61698-9.45779\i$ & $4.42446-7.57498\i$ \\
10& $13.1326-17.9050\i$ & $10.6846-15.9050\i$ & $8.35809-13.4194\i$ & $6.52870-10.9389\i$ & $5.13661-8.77589\i$ \\
11& $14.9197-19.9603\i$ & $12.2295-17.9063\i$ & $9.58311-15.2286\i$ & $7.47640-12.4647\i$ & $5.87560-10.0171\i$ \\
12& $16.7312-22.0133\i$ & $13.8163-19.9208\i$ & $10.8517-17.0713\i$ & $8.45806-14.0308\i$ & $6.63987-11.2955\i$ \\
13& $18.5628-24.0635\i$ & $15.4397-21.9449\i$ & $12.1609-18.9428\i$ & $9.47193-15.6338\i$ & $7.42811-12.6085\i$ \\
14& $20.4109-26.1108\i$ & $17.0953-23.9759\i$ & $13.5077-20.8391\i$ & $10.5165-17.2703\i$ & $8.23918-13.9539\i$ \\
15& $22.2731-28.1552\i$ & $18.7791-26.0118\i$ & $14.8897-22.7565\i$ & $11.5903-18.9376\i$ & $9.07208-15.3295\i$ \\
\hline\hline
\end{tabular}
\caption
    {%
    Dispersive corrections $\omega_n(q)-q$ of the first fifteen helicity $0$ quasinormal mode frequencies,
    for wavenumbers $q = 10$, 20, 40, 80 and 160.
    Both frequencies and wavenumbers are in units of $\pi T$.
    Values were obtained by numerically solving
    eqn.~(\ref{eqn:radial-spin0});
    all digits shown are accurate.
    }
\label{tab:frequencies0}
\end{table}

\section{Numerical techniques}\label{app:numerics}

For solving linear differential equations such as our quasinormal mode equations
(\ref{eqn:radial-eq}), (\ref{eqn:radial-spin1}) and (\ref{eqn:radial-spin0}),
(pseudo)spectral methods are superior to traditional short-range discretization methods.
Spectral methods converge faster, provide greater accuracy for a given
number of discretization points, and allow one to easily enforce boundary conditions
at either end of the computational domain without use of inefficient ``shooting''
techniques.%
\footnote
    {%
    A slightly more detailed discussion of spectral methods
    may be found in ref.~\cite{Chesler:2013lia}.
    For an extensive treatment, ref.~\cite{Boyd:2001} is recommended.
    }
The basic approach, as sketched in section \ref{sec:numerics2},
is to represent the unknown function as a (truncated) expansion in a set of basis
functions, and demand that the original differential equation be satisfied on a
discrete set of points (the ``collocation grid'') within the computational interval.
The optimal grid depends on the choice of basis functions.
When using Chebyshev polynomials up to order $M$, the Chebyshev-Gauss-Lobatto
grid points (\ref{eq:grid}), consisting of the endpoints plus extrema of the highest
order basis function, are an optimal grid.
For functions which are analytic (in a neighborhood of the computational interval),
the Chebyshev expansion converges exponentially rapidly with truncation size $M$.

To solve the helicity $\pm2$ quasinormal mode equation (\ref{eqn:radial-eq}),
for a given numerical value of $q$,
one may directly represent the unknown function $h(u)$ as a Chebyshev sum
(\ref{eq:Chebexp}),
as the desired solution is regular at both $u = 0$ and $u = 1$.
The radial equation (\ref{eqn:radial-eq}) has a regular singular point at each endpoint,
but if the entire equation is multiplied by the $u(1{-}u^4)$ denominator,
then every term remains well-behaved on the $[0,1]$ interval,
including at the endpoints (where the equation effectively becomes first order).
As noted in section \ref{sec:numerics2}, demanding that the resulting equation be
satisfied on each point of the Chebyshev-Gauss-Lobatto collocation grid (\ref{eq:grid})
converts the original differential equation into a finite set of homogeneous linear
equations of the form $M(\omega) \, \vec f = 0$.
The $(M{+}1)\times(M{+}1)$ coefficient matrix $M(\omega)$ is a linear function
of the unknown frequency $\omega$, so the
determinant $\det M(\omega)$ is an $(M{+}1)$-order polynomial in $\omega$.
Constructing this characteristic equation directly,
by evaluating $\det M(\omega)$ for unknown (symbolic) values of $\omega$,
is not an effective computational strategy.
But the linear equation may be trivially recast as a generalized eigenvalue equation
of the form $A \vec f = \omega \, B \, \vec f$, where $A$ and $B$ are purely
numerical matrices.
Such generalized eigenvalue problems may be solved efficiently in $\O(M^3)$ time.

The smallest eigenvalues (in absolute value) converge most rapidly as
the truncation size $M$ increases, with any given eigenvalue
$\omega_n(q;M)$
showing exponential convergence for sufficiently large $M$.
For any given value of $M$, the largest eigenvalues will always be
sensitive to the truncation and hence dominated by discretization artifacts;
at most some fraction of the smallest eigenvalues will be well converged.
As the chosen value of the wavevector $q$ increases, even the lowest
quasinormal mode eigenfunction becomes highly oscillatory, and this
necessitates the use of a truncation size $M$ which grows linearly with $q$.
For sufficiently large $M$,
use of extended precision is also necessary
to avoid excessive round-off error.
For these reasons, a direct numerical solution of the quasinormal mode
equation (\ref{eqn:radial-eq}) becomes quite challenging for values of $q$
beyond about 1000.

Such large-$q$ computational difficulties are not present in the
$q$-independent matching equation (\ref{eqn:hh-matching3}) which
emerged from the WKB analysis of section \ref{sec:subleading-q}.
However, this equation needs to be solved on the positive halfline,
and the equation has an irregular singular point at infinity
plus a regular singular point at the origin.
To find numerical solutions one may work either on the
original halfline $y \in \mathbb R^+$, or on the rotated halfline
(\ref{eq:rotate}) where $y = e^{i \pi/3} \, w$ with $w \in \mathbb R^+$.
To be definite, we describe here our approach when working with the
original form (\ref{eqn:hh-matching3}).

Solutions of interest have
an essential singularity at infinity,
$\bar h(y) \sim y^{-1/4} \exp[ \frac i3 y^{3/2} -i s_\alpha^\infty y^{-1/2}]$,
and $\O(y^{3/2})$ power-law behavior as $y \to 0$.
To apply pseudo-spectral methods to eq.~(\ref{eqn:hh-matching3}), 
we first make a function redefinition which strips off the leading large-$y$ behavior,
\begin{equation}
    \bar h(y) =
    y^{-1/4} \, e^{\frac{i}{3}y^{3/2}} \, H(y) \,.
\end{equation}
The redefined function $H(y)$ satisfies
\begin{align}
    H''
    + (i y^{1/2} - \half y^{-1}) \, H'
    + \half (s_\alpha^\infty \, y^{-1} -\tfrac 7{8} y^{-2}) \, H
    = 0 \,,
\end{align}
and now remains finite and non-zero as $y \to \infty$.
We then map the positive halfline to the computationally convenient finite
interval $[0,1]$ by introducing
\begin{equation}
    u \equiv \big[1+y^{-1/2} \, \big]^{-1} \,,
\end{equation}
or
$
    y = u^2 (1{-}u)^{-2}
$,
and simultaneously extract the leading small $y$ behavior by defining
\begin{equation}
    \tilde H(u) \equiv u^{-3} H(y(u)) \,.
\end{equation}
After writing the resulting equation in a form where all terms remain finite
at $u = 0$ and 1, we arrive at
\begin{align}
    u^2 (1{-}u)^4 \tilde H''
    + 4 u \big[(1{-}u)^3 (1{-}2u) + \tfrac i2 u^3 \big] \tilde H'
    - \big\{(1{-}u)^2[12 u(1{-}u)+\tfrac 74 ] - 2 u^2 (s_\alpha^\infty + 3 i u) \big\}
	\tilde H
    = 0 \,.
\label{eq:final-q-indep}
\end{align}
Solutions of interest to eq.~(\ref{eq:final-q-indep}) are now
regular at both $u = 0$ and 1.
Applying the same pseudo-spectral approximation scheme described above
converts the differential equation to a generalized eigenvalue problem
(with $s_\alpha^\infty$ now the eigenvalue of interest).
Before doing so, however, we make one final variable transformation, setting $u = v^2$
and using a Chebyshev-Gauss-Lobatto grid in $v$, as this was found to improve
convergence of the spectral approximation.
To obtain the results shown in table~\ref{tab:s1-values}, accurate to more
than 12 digits, truncations up to $M = 600$ were used.%
\footnote
    {%
    Using the same strategy,
    convergence of the spectral approximation is even better when working
    with the real form (\ref{eq:rotateeq}) on the rotated halfline.
    Roughly half as many grid points suffice for a given accuracy.
    }

\section{Transformation to infalling coordinates}\label{app:transform}

With the Fefferman-Graham form of the metric (\ref{eq:ds2FG}),
the gauge invariant helicity $\pm1$ combination $Z_1$,
defined in eq.~(\ref{eq:Z1}),
satisfies the equation [\emph{c.f.} (4.26) of ref.~\cite{Kovtun:2005ev}],
\begin{align}\label{eqn:spin-1}
    Z_1''
    -\left[\frac{1}{z} - \frac{\omega^2 f'}{f(\omega^2 - fq^2)}\right] Z_1'
    +\left[ \frac{\omega^2- fq^2}{4z f^2} \right] Z_1
    &=0 \,,
\end{align}
where $f(z) \equiv 1{-}z^2$,
while the gauge invariant helicity 0 combination $Z_2$,
defined in eq.~(\ref{eq:Z2}), satisfies 
\begin{align}\label{eqn:spin-0}
    Z_2''
    -\left[\frac{1+z^2}{z f}+\frac{4q^2z}{q^2(z^2{-}3)+3\omega^2}\right] Z_2'
    + \frac 1f 
    \left[\frac{\omega^2-fq^2}{4z f}-\frac{4q^2 z^2}{q^2(z^2{-}3)+3\omega^2}\right]Z_2
    &=0 \,.
\end{align}
The Fefferman-Graham incoming boundary condition at the horizon,
$Z_i(z) \sim (1{-}z)^{-i \omega/4}$ as $z \to 1$,
can be changed into one of regularity by transforming to infalling coordinates via
\begin{equation}
    z=u^2 \,,\qquad
    \tau=t+\half (\tan^{-1} u+\tanh^{-1} u) \,. 
\end{equation}
This converts the metric (\ref{eq:ds2FG}) into the infalling form (\ref{eqn:metric})
(with $m$ set to unity).
It is convenient to introduce transformed gauge invariant perturbations $\widetilde Z_i$
($i = 1,2$)
such that
\begin{equation}
    e^{i(qx_3-\omega \tau)} \, Z_i =e^{i(qx_3-\omega t)} u^4 \, \widetilde Z_i \,.
\end{equation}
We insert the factor $u^4$ so that the appropriate boundary condition on
$\widetilde Z_i$ is simply that it be regular at $u = 0$.
More explicitly, our redefinition is
\begin{equation}
    Z_i(z(u)) \equiv
    \exp\big[\tfrac i2 \, \omega \, (\tan^{-1}\ u+\tanh^{-1} u)\big] u^4 \,
    \widetilde Z_i (u) \,.
\label{eqn:Z1-rescaling}
\end{equation}
This transformation converts eqs.~(\ref{eqn:spin-1}) and (\ref{eqn:spin-0}) into
eqs.~(\ref{eqn:radial-spin1}) and (\ref{eqn:radial-spin0}), respectively.

\section{Large order, large-\texorpdfstring{$\q$}{q} asymptotics}\label{app:WKB}

To construct a WKB approximation for eigenfunctions satisfying
eq.~(\ref{eq:rotateeq}), valid for large $\lambda$, it is convenient
to rescale the coordinate $w$ by a factor of $\sqrt\lambda$.
If $\kappa \equiv \sqrt\lambda$ and $v \equiv w/\kappa$, then
$f(v) \equiv \h(\kappa v)$ satisfies
\begin{equation}
    \kappa^{-3} \, f''
    =
    \left[
	\tfrac 14 \, v - v^{-1} + \tfrac 34 \, \kappa^{-3} \, v^{-2}
    \right] f \,.
\label{eq:hh-matching-w}
\end{equation}
The linear term in the ``potential'' on the right dominates for large $v$.
The last term $v^{-2}$ is dominant for small $v$,
but this term is negligible for $\O(1)$ values of $v$.
The first two terms in the potential cancel at the point $v=2$, 
which is a turning point in the WKB analysis.
To construct a consistent approximation on the entire halfline,
one must piece together suitable approximations for the solution in
the near-boundary (NB),
classically allowed (WKB-I),
turning point (TP), and
classically forbidden (WKB-II) regions,
illustrated here:
$$
 \begin{tikzpicture}
  \draw (0,0) -- (7,0);
  
  \draw (0,-0.1) -- (0,0.1);
  \node at (0,-0.3) {\small $v\,{=}\,0$};
  
  \draw (3,-0.1) -- (3,0.1);
  \node at (3,-0.3) {\small $v\,{=}\,2$};
  
  \node at (7,-0.3) {\small $v\,{=}\,\infty$};
  
  \draw (0.4,0.3) -- (2.8,0.3);
  \node at (1.6,0.5) {\footnotesize WKB-I};
  
  \draw (2.6,0.6) -- (3.4,0.6);
  \node at (3,0.8) {\footnotesize TP};
  
  \draw (0,0.6) -- (0.8,0.6);
  \node at (0.4,0.8) {\footnotesize NB};
  
  \draw (3.2,0.3) -- (7,0.3);
  \node at (5.1,0.5) {\footnotesize WKB-II};
 \end{tikzpicture}
$$
In the near-boundary (NB) region, $v \ll 1$, the linear
term in the potential is negligible and
(at the order of approximation we are interested in)
may be entirely neglected.
The resulting equation,
$
    \kappa^{-3} \, f''
    =
    \left[
	\tfrac 34 \, \kappa^{-3} \, v^{-2} - v^{-1}
    \right] f
$
has solutions proportional to order-2 Bessel functions.
Only the regular solution,
\begin{equation}
    f_\mathrm{NB}(v) \equiv \sqrt{v} \, J_2(2 \kappa^{3/2} \sqrt{v}) \,,
\end{equation}
satisfies the required boundary condition that the solution vanish as
$\O(v^{3/2})$ as $v \to 0$.
For $v \gg \kappa^{-3}$, this solution (up to an irrelevant overall constant)
behaves as
\begin{equation}\label{eq:hfrak-nb-large}
    f_\mathrm{NB}(v)
   \sim
   v^{1/4}\cos\big(2 \kappa^{3/2} \sqrt{v}-\tfrac{5}{4}\pi \big) \,.
\end{equation}

A WKB ansatz of the usual form,
$
    f_\mathrm{WKB} = \exp\left[\kappa^{3/2}S_0+S_1+\cdots\right]
$,
is applicable in the classically allowed WKB-I region where
$\kappa^{-3} \ll v < 2$ with $2{-}v \gg \kappa^{-1}$.
This ansatz generates a consistent expansion in powers of $\kappa^{-3/2}$.
At next-to-leading order only the first two terms in the potential
contribute, and one finds the oscillatory solutions,
\begin{equation}
    f^\pm(v)
    \equiv
    \left(v^{-1} - \tfrac 14 \, v\right)^{-1/4} \>
    \exp\left[\pm i \, \kappa^{3/2} \int_0^v dv' \>
    \sqrt{v'^{-1} - \tfrac 14 \, v'}\right] .
\label{eqn:hh-WKB-I}
\end{equation}
(Setting to zero the lower limit of integration is a convenient choice
for this arbitrary constant.)
The domain of validity of this solution overlaps that of the near-boundary
approximation when $\kappa^{-3} \ll v \ll 1$.
The linear combination of the two solutions $f^\pm$ which
smoothly matches to the near-boundary solution is
\begin{equation}
    f_\mathrm{WKB-I}(v)
    =
    \left(v^{-1} - \tfrac 14 \, v\right)^{-1/4} \>
    \cos\left[ \kappa^{3/2}
	\left( \int_{0}^v dv' \>
	\sqrt{v'^{-1} - \tfrac 14 \, v'}
	\right)
	- \tfrac 54 \pi
    \right] .
\label{eq:WKBI-sol}
\end{equation}
As $v$ approaches the turning point at 2 (from below),
this solution behaves as
\begin{equation}
    f_\mathrm{WKB-I}(v)
    \sim
    (2{-}v)^{-1/4} \>
    \cos\left\{ \kappa^{3/2}
	\left[ I - \tfrac{\sqrt 2}{3} \, (2{-}v)^{3/2} \right]
	- \tfrac 54 \pi
	\right\} ,
\label{eq:WKBI-tp}
\end{equation}
where
\begin{equation}
    I \equiv
    \int_{0}^2 dv \> \sqrt{v^{-1} - \tfrac 14 \, v}
    =
    \sqrt{2\pi} \> {\Gamma(\tfrac 54)} \bigm/ {\Gamma(\tfrac 74)} \,.
\label{eq:I}
\end{equation}

In the classically forbidden region,
$
    v{-}2 \gg \kappa^{-1}
$,
there are
exponentially growing and decaying solutions.
We require the exponentially decaying solution which behaves as
$
    f(v) \sim
    v^{-1/4} \exp[-\tfrac 13 (\kappa v)^{3/2} (1 + 6 v^{-2}) ]
$
when $v \to \infty$.
The next-to-leading order WKB approximation to this solution is
\begin{equation}
    f_\mathrm{WKB-II}(v)
    =
    \left(\tfrac 14 \, v - v^{-1}\right)^{-1/4} \>
    \exp\left[- \kappa^{3/2} \int_2^v dv' \>
    \sqrt{\tfrac 14 \, v' - v'^{-1}}\right] ,
\label{eq:WKBII-sol}
\end{equation}
where we have again made a convenient choice for the lower limit of integration.
As $v$ approaches the turning point at 2 (from above),
this solution behaves as
\begin{equation}
    f_\mathrm{WKB-II}(v)
    \sim
    (v{-}2)^{-1/4} \>
    \exp\left[
	-\tfrac {\sqrt 2}3 \, \kappa^{3/2} \, (v{-}2)^{3/2} 
	\right] .
\label{eq:WKBII-tp}
\end{equation}

The remaining task is to connect the WKB solutions (\ref{eq:WKBI-sol}) and
(\ref{eq:WKBII-sol}) across the turning point at $v=2$.
Within the turning point region, $ |v{-}2| \ll 1 $,
the potential may be linearized about $v = 2$ and,
at the order of interest,
the $\kappa^{-3} v^{-2}$ term in the potential may be neglected.
This gives
$
    \kappa^{-3} \, f'' = \half (v{-}2) \, f
$,
whose solutions are Airy functions.
Only the Airy function of the first kind can match onto the decaying WKB-II
solution at large $v$,
so the solution within the turning point region is
\begin{equation}
    f_\mathrm{TP}(v)
    =
    \mathrm{Ai}\big(\kappa (v{-}2)/2^{1/3}\big) \,.
\label{eq:h-tp}
\end{equation}
For $v{-}2 \gg \kappa^{-1}$, the asymptotic behavior of this Airy function
coincides with the near turning point behavior (\ref{eq:WKBII-tp})
of the WKB-II solution, as required.
On the other side of the turning point,
when $2{-}v \gg \kappa^{-1}$,
the asymptotic behavior of the Airy function with large negative argument
gives
\begin{equation}
    f_\mathrm{TP}(v)
    \sim
    (2{-}v)^{-1/4} \,
    \cos\left[
	\tfrac {\sqrt 2}3 \, \kappa^{3/2} \, (2{-}v)^{3/2}
	- \tfrac 14 \pi
    \right] .
\end{equation}
This agrees with the oscillatory behavior (\ref{eq:WKBI-tp})
of the WKB-I solution near the turning point,
up to a shift in the phase.
For a consistent solution, these phase shifts must also agree
modulo $\pi$
(since a difference of $\pi$ can be absorbed by a sign flip of an
overall coefficient).
Consequently, we require that
\begin{equation}
    \kappa^{3/2} \, I = (n + \half) \, \pi  \,,
\end{equation}
for some integer $n$.
Solving for the eigenvalue $\lambda = \kappa^2$ and inserting the
value (\ref{eq:I}) of $I$ yields the next-to-leading approximation
for large order eigenvalues,
\begin{equation}
    \lambda_n^{\rm WKB}
    =
    \left[
	(n + \half) \sqrt{\tfrac \pi2} 
	\> {\Gamma(\tfrac 74)} \bigm/ {\Gamma(\tfrac 54)}
    \right]^{4/3}
    \,.
\label{eq:lambdan}
\end{equation}
Inclusion of higher order terms in the WKB expansion will generate
relative corrections to this result of order $\kappa^{-3} \sim n^{-2}$.
One may verify that the allowed region solution (\ref{eq:WKBI-sol})
has $n{-}1$ nodes when $\lambda = \lambda_n^{\rm WKB}$
implying that, as written, $n$ is the level number
when counting starts from 1.
Recalling [from eq.~(\ref{eq:rotateeq})] that the eigenvalue
$\lambda = \half s_\alpha^\infty \, e^{i \pi/3} = \half c_n$
one finds the result (\ref{eqn:s-alpha-asympt}) quoted earlier
for the large order behavior of the asymptotic coefficients $\{c_n\}$.

\goodbreak

\bibliography{qnm}
\end{document}